\documentclass[a4paper,12pt]{article}
\usepackage{amsmath,amssymb}
\usepackage{hyperref}
\usepackage{graphicx}
\usepackage{epstopdf}
\usepackage{mathrsfs}
\usepackage{float}
\usepackage[normalem]{ulem}  

\newcommand{\be}{\begin{equation}}
\newcommand{\ee}{\end{equation}}
\newcommand{\f}{\frac}
\newcommand{\s}{\sqrt}
\newcommand{\p}{\partial}

\newcommand{\bea}{\begin{eqnarray}}
\newcommand{\eea}{\end{eqnarray}}
\newcommand{\ba}{\begin{align}}
\newcommand{\ea}{\end{align}}

\newcommand{\lsim}{\mathrel{\hbox{\rlap{\lower .55ex
\hbox{$\sim$}} \kern-.3em \raise.4ex \hbox{$<$}}}}
\newcommand{\gsim}{\mathrel{\hbox{\rlap{\lower.55ex
\hbox{$\sim$}} \kern-.3em \raise.4ex \hbox{$<$}}}}

\begin{document}


\def\gap#1{\vspace{#1 ex}}
\def\be{\begin{equation}}
\def\ee{\end{equation}}
\def\bal{\begin{array}{l}}
\def\ba#1{\begin{array}{#1}}  
\def\ea{\end{array}}
\def\bea{\begin{eqnarray}}
\def\eea{\end{eqnarray}}
\def\beas{\begin{eqnarray*}}
\def\eeas{\end{eqnarray*}}
\def\del{\partial}
\def\eq#1{(\ref{#1})}
\def\fig#1{Fig \ref{#1}} 
\def\re#1{{\bf #1}}
\def\bull{$\bullet$}
\def\nn{\\\nonumber}
\def\ub{\underbar}
\def\nl{\hfill\break}
\def\bibi{\bibitem}
\def\ket{\rangle}
\def\bra{\langle}
\def\vev#1{\langle #1 \rangle} 
\def\mattwo#1#2#3#4{\left(\begin{array}{cc}#1&#2\\#3&#4\end{array}\right)} 
\def\tgen#1{T^{#1}}
\def\half{\frac12}
\def\floor#1{{\lfloor #1 \rfloor}}
\def\ceil#1{{\lceil #1 \rceil}}

\def\mysec#1{\gap1\ni{\bf #1}\gap1}
\def\mycap#1{\begin{quote}{\footnotesize #1}\end{quote}}

\def\Om{\Omega}
\def\a{\alpha}
\def\b{\beta}
\def\l{\lambda}
\def\g{\gamma}

\def\lan{\langle}
\def\ran{\rangle}

\def\bit{\begin{item}}
\def\eit{\end{item}}
\def\benu{\begin{enumerate}}
\def\eenu{\end{enumerate}}

\def\tr{{\rm tr}}


\def\bT{\bar T}
\def\cO{{\mathcal O}}
\def\bl{{\bf L}}

\begin{titlepage}
\begin{flushright}
TIFR/TH/16-16
\end{flushright}

\vspace{.4cm}
\begin{center}
\noindent{\Large \textbf{Finite size effect on dynamical entanglement entropy:
CFT and holography}}\\
\vspace{1cm}
Gautam Mandal$^{a,}$\footnote{e-mail:\; mandal@theory.tifr.res.in},
\;
Ritam Sinha$^{a,}$\footnote{sinha@theory.tifr.res.in}, and 
\;
 Tomonori Ugajin$^{b,}$\footnote{ugajin@kitp.ucsb.edu}

\vspace{.5cm}
\centerline{{\it $^{a}$Department of Theoretical Physics}}
\centerline{{\it Tata Institute of Fundamental Research, Mumbai 400005, India.} }
\vspace{.5cm}
  {\it
 $^{b}$Kavli Institute for Theoretical Physics, University of California, \\
Santa Barbara, 
CA 93106, USA\\
\vspace{0.2cm}
 }
\end{center}


\begin{abstract}

Time-dependent entanglement entropy (EE) is computed for a single
interval in two-dimensional conformal theories from a quenched initial
state in the presence of spatial boundaries.  The EE is found to be
periodic in time with periodicity equal to the system size $L$. For
large enough $L$, the EE shows a rise to a thermal value
(characterized by a temperature $1/\b$ determined by the initial
state), followed by periodic returns to the original value. This works
irrespective of whether the conformal field theory (CFT) is rational
or irrational. For conformal field theories with a holographic dual,
the large $c$ limit plays an essential role in ensuring that the EE
computed from the CFT is universal (independent of the details of the
CFT and of boundary conditions) and is exactly matched by the
holographic EE. The dual geometry is computed and it interpolates
between a BTZ black hole at large $L$ and global AdS at large $\b$.

\end{abstract}


\end{titlepage}

\tableofcontents

\section{Introduction and Summary}


Quantum quench refers to non-equilibrium dynamics caused by a sudden
change of the Hamiltonian. Interest in studies of this phenomenon has
partly been spurred by its experimental realization in cold atom
systems. The pioneering article \cite{Greiner:2002} by Greiner et al
addressed the question of a quantum phase transition following a
quantum quench in a system of ultracold atoms. Theoretical interest
has also grown in this area following seminal work on quantum quenches
involving conformal and integrable models, especially in two
dimensions \cite{Rigol:2007a, Calabrese:2012}. An important question
addressed in these studies is whether a closed dynamical system,
following a quantum quench, approaches a thermal equilibrium.  For
conformal field theories with a holographic dual, this question takes
an independently interesting meaning, since the holographic dual of
thermalization turns out to be gravitational collapse to a black hole
\cite{Bhattacharyya:2009uu,Chesler:2008hg,Balasubramanian:2011ur}.

In two dimensions, with infinite spatial size, quantum quench to a
critical point has been discussed extensively by Calabrese and Cardy
in a series of papers, starting with \cite{Calabrese:2004eu,
  Calabrese:2005in}. The final hamiltonian in this problem corresponds
to a 2D CFT and the quenched state is effectively replaced, at least
for the physics of long wavelengths, by a simpler choice of initial
wavefunction which is a conformal boundary state with a UV cutoff
scale. Such states are translationally invariant, and the quench is
called a global quench. The post-quench time evolution of various
observables is given by path integrals on a strip whose width is
determined by the cut-off scale. Local observables confined within an
interval can be shown to thermalize at rates with universal properties
\cite{Cardy:2014rqa}. This happens even in the presence of arbitrary
number of conserved charges, e.g., for integrable CFTs where the
thermal ensemble is given by a generalized Gibbs ensemble (GGE)
\cite{Mandal:2015jla, Cardy:2015xaa}. In a holographic context, the
thermalization in the CFT matches the decay of quasinormal modes
\cite{Horowitz:1999jd} of black holes (in case of the GGE, it
corresponds to quasinormal modes of the higher spin black hole
\cite{Mandal:2015jla}). Another important time-dependent observable in
the CFT is the entanglement entropy (EE); this was first calculated in
the article \cite{Calabrese:2005in} in the quench model of
\cite{Calabrese:2004eu} where the growth of entanglement was also
given an interpretation in terms of quasiparticles moving at the speed
of light (for a recent work on a bound on the spread of entanglement,
see \cite{Casini:2015zua}). In a holographic context, the
time-dependent EE of \cite{Calabrese:2005in} has been matched with
computations in black hole geometries \cite{Hartman:2013qma} (see also
\cite{Caputa:2013eka}).

The above discussions have been generalized to introduce
inhomogeneities in the initial state by applying a conformal
transformation, in which case time evolution of various observables
can be understood in terms of path integrals on appropriate
inhomogeneous strips \cite{Sotiriadis:2008}. A holographic dual of
this set up was studied in \cite{Ugajin:2013xxa}, and in a related
context involving the eternal black hole, in
\cite{Mandal:2014wfa}. Other work in this regard includes studies of
local quenches in various contexts, see, e.g. \cite{Asplund:2014coa,
  Caputa:2014vaa, Asplund:2013zba, Astaneh:2014fga}.

In much of the above discussion, the spatial extent of the system is
infinite (or much larger than any other length scale in the
problem). The effect of a finite spatial extent on thermalization has
been a topic of long interest, e.g. in the context of the FPU problem
\cite{FPU-1955,FPU-scholarpedia,FPU-scholarpedia2}. From the point of
view of experimental realizations, in case of 2D CFT's, the spatial
boundaries can be thought of as representing impurities in a quantum
critical system.  Thermalization, or lack thereof, for global quenches
in a finite system of length $L$ has recently been studied in
\cite{Cardy:2014rqa}, where it has been shown that for rational CFT's,
the quenched state returns to itself after a finite period of time
which is a multiple of $L/2$. The reason for this `revival' was
roughly that the time-dependence of the wavefunction, for rational
highest weights, is a sum of terms with rational periods. Correlation
functions and entanglement entropy of subsystems also follow a
periodic behaviour.  The periodic behaviour of entanglement entropy
was explicitly confirmed in the global quench on a circle for free
fermions in \cite{Takayanagi:2010wp}. In \cite{Kuns:2014zka}, the
periodic behaviour observed in \cite{Cardy:2014rqa} is explained by
interpreting the quenched state in terms of a Lorentzian-signature
conformal transformation of the ground state on the strip; this paper
also comments on a holographic formulation.  For other discussions on
quenches on the segment, we refer to \cite{Guo:2015uwa,
  daSilva:2014zva, Engelhardt:2015axa} and the recent discussion
\cite{deBoer:2016bov}. Revival in higher dimensional field theories
has been considered in \cite{Cardy:2016lei}; holographic entanglement
entropy has also been discussed in 2+1 dimensional systems with finite
size \cite{Abajo-Arrastia:2014fma} where a partial revival has been
found.


\gap3

The main focus of our paper will be to study the effect of finite
system size on various time-dependent phenomena. Below we summarize
the contents and main results of our paper:

\gap3

(i) We study time evolution of observables in a 2-dimensional CFT
starting from a quenched state with a UV cutoff scale $1/\b$ (as
described above) in the presence of finite spatial boundaries (with
system size $L$). The tool used in the CFT computations is a conformal
map (the Christoffel-Schwarz transformation), as in
\cite{Kuns:2014zka}, which converts a rectangular geometry to an upper
half plane (see Section \ref{sec:map} and Appendix \ref{app:map}). The
map gets simplified in the limits of large $L$ and large $\b$
respectively and can be identified with known versions of the map from
a cylinder to a plane.

(ii) We show that the spatial boundaries lead to locally thermalized
regions (characterized by a {\it temperature} $1/\b$) which merge and
split periodically at regular time intervals given by the system
size. We also quantitatively compute thermalization of certain
observables on intermediate time scales prior to the occurrence of the
`revival' phenomena discussed above. We show that in the limit of
large $L$, the relaxation rate agrees with the known results for
infinite systems (see Sections \ref{sec:one-pt}, \ref{sec:energy},
\ref{sec:one-pt-x-t} and Appendix \ref{app:one-pt}). Part of the above
discussion has already appeared in
\cite{Kuns:2014zka,Engelhardt:2015axa} (see also
\cite{deBoer:2016bov}).

(iii) In Section \ref{sec:bulk} we present a bulk dual of the CFT on a
rectangle, following the AdS/CFT proposal for CFT with boundaries
\cite{Karch:2000ct, DeWolfe:2001pq, Takayanagi:2011zk, Fujita:2011fp,
  Nozaki:2012qd}, coupled with \cite{Roberts:2012aq, Ugajin:2013xxa,
  Mandal:2014wfa} which discuss a {\it large diffeomorphism} (see
Eqn. \eq{roberts}) that reduce to the (analytically continued)
conformal map discussed above.  The periodicity of various observables
mentioned above can be interpreted in terms of the time-periodicity of
the above large diffeomorphism, which effects a time-periodic change
of the (regulated) AdS boundary and, in turn, causes such changes in
geometric quantities such as lengths of geodesics tied to the
boundary. 
 
(iv) The main results of our paper, presented in Sections
\ref{sec:EE}, \ref{sec:bulk} and \ref{sec:quasiparticle}, concern the
computation of the entanglement entropy (EE) of a single interval of
length $l$.  We use the conformal map described above, and the method
of images, to relate the CFT computation of EE to a four-point
function of `twist operators' on the plane. We also compute the EE
using holographic methods (Section \ref{sec:hEE}), developing on
earlier work by one of the authors \cite{Ugajin:2013xxa}. The
holographic result is universal and does not depend on the specific
CFT (except on the central charge). The CFT result, on the other hand,
involves a four-point function which generically depends on the
specific CFT. At large $L$, the CFT four-point function factorizes,
becomes universal, and readily agrees with the holographic result. The
analysis for general $L/\b$, however, is much more subtle; we show
that (see Section \ref{large-c}), although the CFT four-point function
does not {\it a priori} factorize, it takes a universal form provided
one takes an appropriate large $c$ limit discussed recently
\cite{Hartman:2013mia, Asplund:2014coa, Fitzpatrick:2014vua,
  Leichenauer:2015xra}. We show that this new universal form then
agrees with the result obtained from holography.

(vi) Another novel computation of our paper (see Section
\ref{sec:quasiparticle}) is that of the EE by adapting the
quasiparticle method \cite{Calabrese:2005in} to the presence of
spatial boundaries. The boundaries leads to hard wall reflection of
the quasiparticles, causing periodic entry and exit of EPR partners to
and from the interval of interest. We compute the resulting
entanglement entropy. The result quantitatively agrees with the large
$L$ results discussed in the last paragraph.

(vii) We should remark that in \cite{Cardy:2014rqa}, revival was
investigated by using the fidelity function $|\bra \psi_0 |\exp[-i H
  t]\psi_0 \ket|^2$\, \footnote{This is a simpler version of the
  so-called Loschmidt echo.}; it was pointed out that no such revival
was expected in the presence of a continuum of (or more generally,
incommensurate) conformal weights, e.g.  in an irrational conformal
field theory. The periodicity we find in this paper in entanglement
entropy and other observables is observed, however, in {\it any} CFT,
including large $c$ theories with a holographic dual. This is tied to
the fact that these observables are not sufficient to distinguish
between rational and irrational theories; see \cite{deBoer:2016bov}
for a recent discussion.

\section{\label{sec:CFT}CFT with finite spatial boundary}

In this section, we will describe quantum quench in the presence of a
spatial boundary using conformal field theory methods. We will review
some known results \cite{Cardy:2014rqa, Kuns:2014zka} and some new
results for time-dependent one-point functions. We will discuss
CFT computation of EE in the next section.

We will consider spatial boundaries at $x= \pm L/2$, and following
\cite{Calabrese:2005in}, an initial state of the form,
\begin{align}
|\psi_0 \ket = e^{-\b H/4} | B \ket.
\label{psi-0}
\end{align}
where the state $|B \ket$ is a conformal boundary state (the state
$|\psi_0 \ran$ can be regarded as an approximation to a realistic
quench state \cite{Das:2014hqa,Das:2014jna,Mandal:2015kxi}).  The
  parameter $\beta$ can be regarded as a length scale which cuts off
  the UV modes to render the state
  normalizable.\footnote{\label{ftnt:beta}We will find below that for
    large enough $L/\b$, the energy of this state coincides with the
    energy of a thermal ensemble characterized by an inverse
    temperature $\b$ (see \eq{thermal}). Hence, from here on we will
    refer to $1/\b$ as a `temperature', although we should remember
    that we are still dealing with a pure state and the nomenclature
    is only a formal one.}  Following \cite{Calabrese:2005in}, we will
  view this wavefunction as the result of a Euclidean time evolution
  from a boundary state $| B \ket$ at $\tau = -\b/4$ to $| \psi_0
  \ket$ at $\tau=0$. Real time evolution of \eq{psi-0} is described by
  continuing $\tau$ to complex values:
\begin{align}
|\psi(t) \ket= e^{-i H t}|\psi_0 \ket =  e^{-\tau H} | B \ket, \; 
\tau= \b/4 + i t.
\label{psi-t}
\end{align}
We would be interested in time-dependent quantities such as
(a) the equal-time correlators 
\begin{align}
\bra O_1(x_1,t)... O_n(x_n,t) \ket  
& \equiv   \bra \psi(t)| O_1(x_1) ... O_n(x_n)|\psi(t) \ket.
\label{correlators}
\end{align} 
or (b) the entanglement entropy $S_{EE}(t)$ of an interval $A=[-l/2,
  l/2]$ when the system as a whole is described by the wavefunction
\eq{psi-t}. As discussed in \cite{Calabrese:2005in} (see below),
$S_{EE}(t)$ is related to a two-point correlator of the above kind.

As mentioned above, the boundary state $| B \ket$ implements a certain
boundary condition on the time boundary $\tau=-\b/4$.  When the
\emph{same} boundary conditions are also imposed at the spatial
boundaries $x= \pm L/2$, then the correlators \eq{correlators} can be
evaluated by a functional integral over a rectangle (see figure
\ref{fig:map}), with boundary condition relevant to the boundary state
$|B\ket$ imposed on all sides of the rectangle.\footnote{In case the
  spatial boundary conditions are different from the temporal boundary
  conditions, one needs to insert some boundary operators at the
  corners of the rectangle \cite{Cardy:1984bb}.}

\begin{figure}
\begin{center}
\kern-20pt\includegraphics[width=8cm,height=4.5cm]{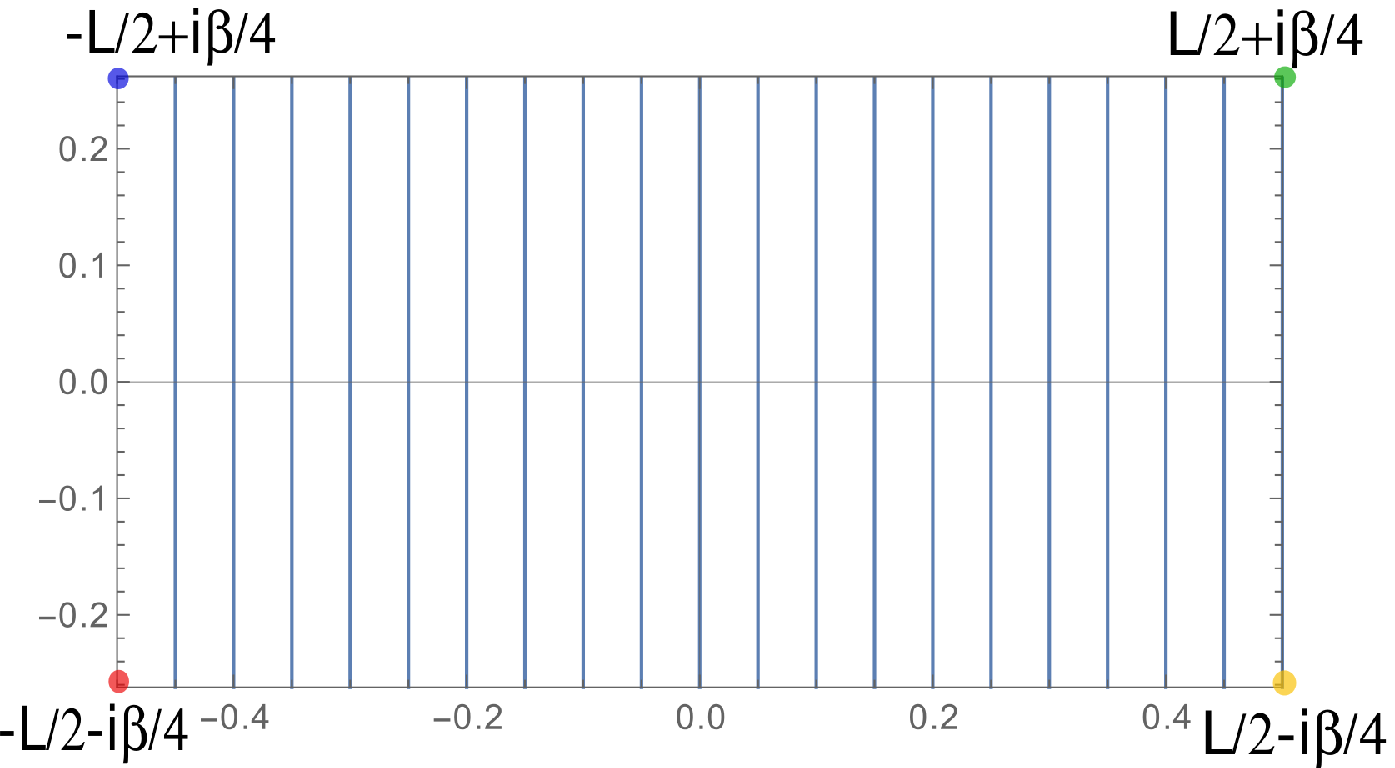}
\kern+20pt\includegraphics[width=7cm,height=7cm]{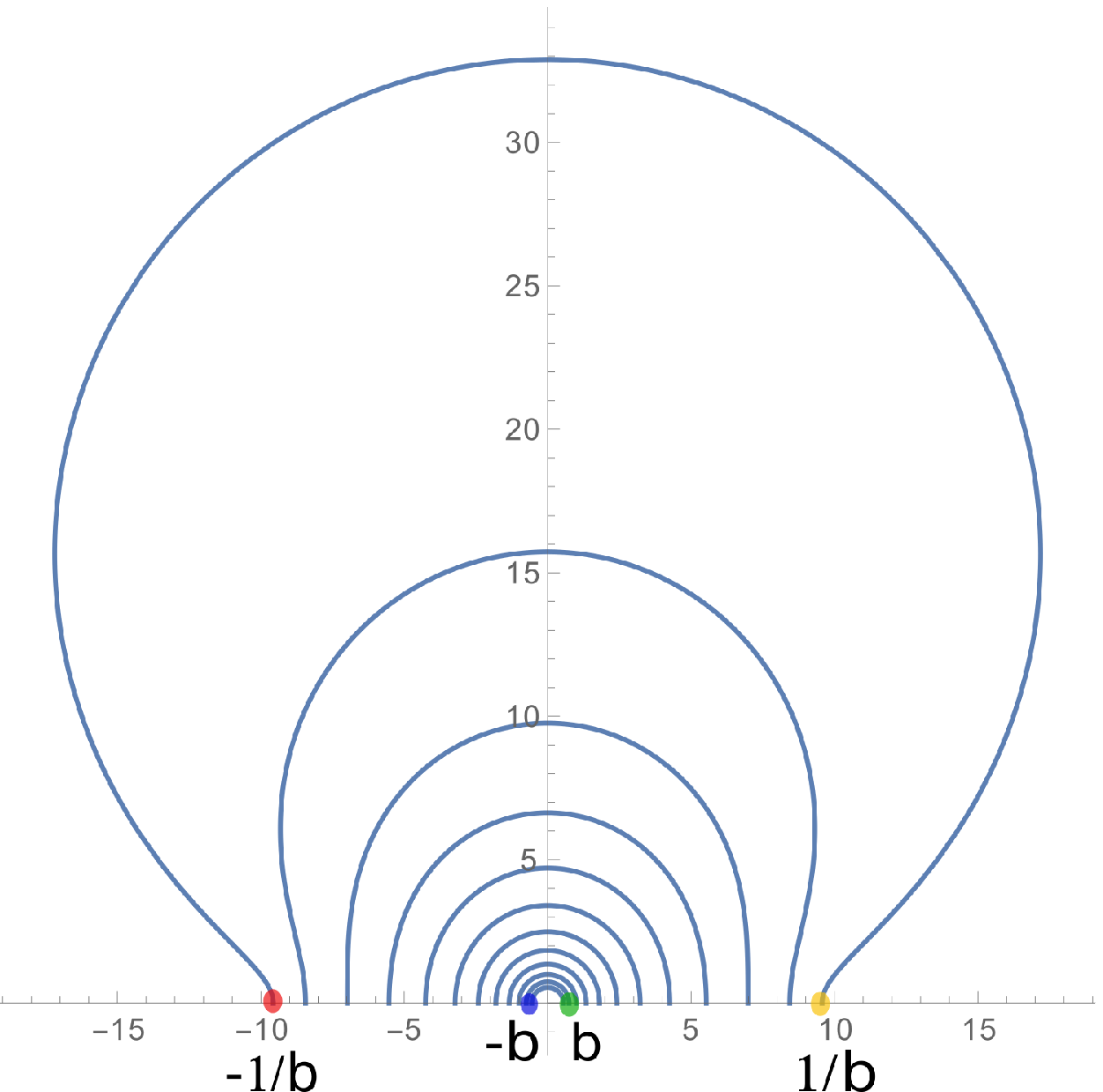}
\end{center}
\caption{\footnotesize  
Maps between the upper half plane and the rectangle. The colour
coding represents the mapping of the corners to the boundary of the
UHP. See (\ref{inverse-map2}) and (\ref{map2}). The time evolution
contours in the rectangle are mapped to the UHP as shown on the
right. We have chosen $L=\beta=1$.}
\label{fig:map}
\end{figure}

\subsection{Mapping the rectangle to the upper half plane}\label{sec:map}

Functional integrals over a rectangular region are not easy to
compute directly.  However, since we are dealing with a CFT, and the
boundary conditions do not break conformal symmetry, the CFT
correlators \eq{correlators} are covariant under conformal
transformations. We can thus reduce the computation to that on the
upper half plane (UHP) by using a conformal map from the rectangle to
the UHP. The necessary map for this purpose is a Christoffel-Schwarz
transformation\footnote{The transformation we use here is a little different from that found
in literature. The reason is that we want to explore both the low and the high 
values of $L/\b$ 
using this map by simply tuning the parameter $b$.}, defined by
\begin{align}
w(z)= A\int^{z}_{0} \f{dz}{\s{(z^2-b^2)(z^2-\f{1}{b^2})}} +B,
\label{christoffel}
\end{align}
Here, $z$ is a complex coordinate on the UHP (defined by the region
Im$(z) \ge 0$ of the plane), whereas $w= x + i \tau$ is a complex
coordinate parametrizing the above-mentioned rectangular region
(bounded by the lines $|{\rm Re}(w)| \le L/2$, $|{\rm Im}(w)| \le
\b/4$, see figure \ref{fig:map}). $A, B$ are constants which determine
the images of the corners of the rectangle on the boundary of the UHP.
Without loss of generality, the four image points can be chosen to
have $z$-coordinates: $(z_1,z_2,z_3,z_4)=(-1/b, -b, b, 1/b)$ with $(0
\le b \le 1)$.  We will look for a map which sends these points to the
following corners of the rectangle respectively:
\begin{align}
\kern-10pt w\left(-\frac 1 b\right)=- \frac L 2-i \frac \b 4,\, 
w(-b)= -\frac
L 2+i \frac \b 4,\, w(b)= 
\frac L 2 + i \frac \b 4 ,\, w\left(\frac 1 b\right)=
\frac L 2 - i \frac \b 4.
\label{images-2}
\end{align}
The required map is discussed in  \cite{Kuns:2014zka} 
(see also Appendix \ref{app:map}). For convenience,
we reproduce here the map $z(w)$ \eq{inverse-map2} 
from the rectangle to the UHP
\begin{align}
z(w)=b\; {\rm sn} \left[ \f{4 K(b^4)}{i \beta} \left(w-\f{L}{2} \right),b^4 
\right],\;\; \bar z(\bar w)
=b\; {\rm sn} \left[ \f{4 K(b^4)}{-i \beta} 
\left(\bar w-\f{L}{2} \right),b^4 \right]
\label{inv-map2}.
\end{align}
The parameter $b$ determines the aspect ratio of the rectangle:
\begin{align}
\f{\beta}{L} =\f{4K(b^4)}{K(1-b^4)} .
\end{align}
We show in Appendix \ref{app:map} that the map \eq{inv-map2} can also
be regarded as a map from the torus (a product of two circles of sizes
$2L, \b$) owing to the periodicity properties \eq{periods}. The map
also has a large $L$ limit \eq{high-limit-app}
\begin{align}
z(w)= i \exp[-2\pi w/\b],
\label{high-limit}
\end{align}
and a low temperature\footnote{\label{ftnt:beta2}We use the word
  `temperature' to refer to $1/\beta$ in the sense of footnote
  \ref{ftnt:beta}.}  limit \eq{low-limit-app}
\begin{align}
z(w)=i \cot \left(\frac\pi 4 + \f{\pi w}{2 L}\right).
\label{low-limit}
\end{align}

\subsection{\label{sec:one-pt}One Point Function}


In this section, we will evaluate a one-point function $\lan O(x,t)
\ran$ as defined in \eq{correlators} (part of the above discussion has
already appeared in \cite{Kuns:2014zka,Engelhardt:2015axa}, see also
\cite{deBoer:2016bov}).  The evaluation would include calculating the
one point function in Euclidean time $\tau$ and then analytically
continuing to Lorentzian time $t$. We shall however, first, use the
map $z(w)$ \eq{inv-map2} to relate the one-point function on the
rectangle to that on the UHP.

\paragraph{Primary operators}

For a primary operator $O(w, \bar w)$, of dimension $(h, \bar h)$,
the one-point function on the rectangle becomes
\begin{equation}
 \langle O(w,\bar w)\rangle_{\text{rect}}=\bigg(\f{\partial
   z}{\partial w}\bigg)^h\bigg(\f{\partial{\bar z}}{\partial{\bar
     w}}\bigg)^{\bar h} \langle O(z,\bar
 z)\rangle_{\text{UHP}}.
\label{conf-primary}
\end{equation}
For a holomorphic operator, or an operator with $\bar h \ne h$, the
above one-point function on the UHP vanishes. However, for a primary
operator of the form $O_{h, h}(w, \bar w) = \phi_h(w)
\phi^\dagger_h(\bar w)$, (with $\bar h=h$)
\footnote{Here we allow for complex operators $\phi$; $\phi^{\dagger}$
  denotes the hermitian conjugate.} the above one-point function on
the UHP is non-zero and is given by the `method of images'
\cite{Cardy:1984bb}.  According to this method, the conformal boundary
condition on the UHP amounts to replacing the antiholomorphic operator
$\phi^{\dagger}_h(\bar z)$ from the point $P=(z,\bar z)$ by a
holomorphic operator $\phi_h(z')$ at the image point $P'=(z', \bar
z')$ (with $z'= \bar z, \bar z'=z$). The one-point function is now
given by the holomorphic 2-point function on the plane\footnote{Up to
  a constant $A_b$ which depends on the operator $O$ and the boundary
  state $|B\ket$. We will assume that $A_b \ne 0$; the precise value
  of this constant will be unimportant and we will drop it
  henceforth.}
\[
\lan O_{h,h}(P) \ran_{UHP}=A_b \lan  \phi_h(P) \phi^{\dagger}_h(P') \ran_{\mathbb{C}}
= A_b  \lan \phi_h(z) \phi^{\dagger}_h(z') \ran_{\mathbb{C}}
\]
Hence, the original one-point function, for primary operators
$O_{h,h}(w,\bar w) \equiv \phi_h(w) \phi_h^{\dagger}(\bar w)$ reduces to
\begin{equation}
 \langle O_{h,h}(w,\bar w)\rangle_{\text{rect}}=\bigg(\f{\partial
   z}{\partial w}\bigg)^h\bigg(\f{\partial{\bar z}}{\partial{\bar
     w}}\bigg)^{h} 
\lan \phi_h(z) \phi^{\dagger}_h(z') \ran_{\mathbb{C}}
\label{connected-2pt-primary}
\end{equation}

\paragraph{Quasiprimary operators}

In case the operator $O(w, \bar w)$ is quasiprimary, it mixes with
lower dimension operators under conformal transformations, leading to
additional terms in \eq{conf-primary}. For example, for
the holomorphic stress tensor  $T(w) \equiv T_{ww}(w)$, the 
conformal transformation to the UHP is given by
\begin{align}
 \langle T(w)\rangle_{\text{rect}}=\bigg(\f{\partial
   z}{\partial w}\bigg)^2 \langle T(z)\rangle_{\text{UHP}}-\f{c}{12} \{z, w \}
= -\f{c}{12} \{z, w \}
\label{Tww}
\end{align}
In the second equality we have used the fact that $\lan T(z) \ran$ on
the UHP is the same as that on the plane (since it does not have an
antiholomorphic factor), and hence vanishes. The last expression
contains the Schwarzian derivative
\begin{align}
\{z, w \}= \f{2(\del_w^3 z)(\del_w z)-3(\del_w^2 z)^2}{2(\del_w z)^2}
\label{def-sch}
\end{align}
A similar formula is true for the antiholomorphic stress tensor 
$\bar T(\bar w)$.

For the operator $O(w, \bar w)= :T(w) \bar T(\bar w):$, by using
a combination of the above techniques, we get a generalization
of the formula \eq{connected-2pt-primary}:
\begin{align}
 \langle :T(w)\bar T(\bar w):\rangle_{\text{rect}}=\bigg(\f{\partial
   z}{\partial w}\bigg)^2 \bigg(\f{\partial
   \bar z}{\partial \bar w}\bigg)^2\langle T(z) T(z_1) \rangle_{\mathbb{C}}
+ \left|\f{c}{12} \{z, w \}\right|^2
\label{TTbar}
\end{align}

\subsubsection{\label{sec:analytic-1pt}Analytic 
continuation to real time correlators}

As discussed before, the time-dependent wavefunction \eq{psi-t},
or equivalently the time-dependent Heisenberg operators $O(x,t)$
can be realized by analytically continuing $\tau$
to imaginary values $\tau = it$. Thus, $O(w,\bar w)= O(x,\tau)$
can be interpreted as the time-dependent operator $O(x,it)$. In terms
of the coordinates on the rectangle, the analytic 
continuation reads\footnote{Note the convention $x^\pm 
\equiv x\mp t$.}  
\begin{align}
\{w, \bar w\}= x\pm i\tau  \xrightarrow{\tau= it}  x\mp t
\equiv x^\pm
\label{analytic}   
\end{align}
This `Wick rotates' the Euclidean rectangle to the Lorentzian
geometry, 
\begin{align}
M_L= {\mathbb{I}} \times R \ni  x_\pm= x \mp t, \; x\in 
\mathbb{I} = [-L/2, L/2],
t\in \mathbb{R}
\label{lorentz-man}
\end{align}
Note that such an analytic continuation is possible since the
Euclidean observables are separately analytic in $(z, \bar z)$, and
hence in $(w, \bar w)$. We will use this rule below to explicitly
compute the time-dependence of one-point functions and later on, the
single interval entanglement entropy.

We note here that although for these applications, we do not need to
have an analytic continuation of the complex $z, \bar z$ plane, we
will indeed find an analytic continuation of the Euclidean map
\eq{inv-map2}, {\it viz.}  \eq{lorentz}, from the the above geometry
$M_L$ to the Lorentzian $\mathbb{R}^2$. The above map is many-to-one
and one can choose a fundamental domain of the map to be a
diamond-shaped region $\mathbb{D}:\{ x_\pm \in (-L/2, L/2)\} \subset
M_L$ in Section \ref{sec:bulk} to build a bulk geometry dual to a CFT
on $M_L$. It has been suggested in \cite{Kuns:2014zka}, and further
elaborated in \cite{deBoer:2016bov}, that thermalization appears to
happen when one confines to this diamond (which is natural from the
viewpoint of the $\mathbb{R}^2$, whereas the actual spacetime is all
of $M_L$, with its built-in recurrence. In Section \ref{sec:bulk}, the
map \eq{lorentz} is crucially used to construct a dual geometry for
the CFT state. We will discuss this map in detail in Appendix
\ref{zplus-minus}.


\subsection{\label{sec:energy}Behavior of the Energy density}

In Euclidean CFT, the energy density is given by
\begin{align}
{{\cal E}}_{Eucl}(w, \bar w)= \lan T_{\tau\tau} \ran
=- \left( \lan T(w) \ran +\lan \bar T(\bar w) \ran  \right) =   
\f{c}{12}\left( \{z, w \} + \{\bar z, \bar w\} \right)
\label{e-euclid}
\end{align}
where in the last step we have used \eq{Tww} and its antiholomorphic
counterpart.  In the limits of high and low temperature, \eq{high-limit} and
\eq{low-limit} respectively, the Schwarzian derivative is easy to compute, leading
to the constant values
\begin{align}
{{\cal E}}_{Eucl}=\left\{ 
\begin{array}{l}
-{c\pi^2}/{3 \beta^2}, \; \;  \b \ll  L\\
+{c \pi^2}/{12 L^2}, \; \;  \b \gg  L
\end{array}
\right.
\end{align}
Using the methods of Section \ref{sec:analytic-1pt} and the analytic continuation\eq{analytic},
the energy density in the Lorentzian theory is,
\begin{align}
{{\cal E}}(x,t)&=
\lan T_{++}(x_+)+ T_{--}(x_-) \ran
=\lan T_{tt} \ran= - \lan T_{\tau\tau} \ran
= - {{\cal E}}_{Eucl}(w=x_+, \bar w=x_-)
\label{e-lorentz}
\end{align}
Now, the high and low temperature limits are,
\begin{align}
{{\cal E}}(x,t) &
=+{c\pi^2}/{3 \beta^2}, \; \;  \b \ll  L \label{thermal}\\
& =-{c \pi^2}/{12 L^2},\; \;  \b \gg  L \label{casimir}.
\end{align}
which can be identified with the well-known expressions for the
thermal energy and Casimir energy respectively. 

We note here that some of the results in this subsection have been
obtained and discussed in \cite{Cardy:2014rqa, Kuns:2014zka}; we
include these here for the sake of completeness.

\subsubsection{Time-dependence}

Away from the above two limits, the energy density \eq{e-lorentz}
is both space and time-dependent. We display the behaviour of
the `normalized' dimensionless quantity
\begin{align}
\widetilde{\cal E}=  \f{{\cal E}L^2}{c}
\label{normal-energy}
\end{align}
in Figures \ref{fig:energy-t} and
\ref{fig:energy-2D}. In Figure \ref{fig:energy-t}, one can clearly see two
crests getting reflected back and forth periodically at the boundary
walls. These correspond to the holomorphic and antiholomorphic stress
tensors respectively.

\begin{figure}[H]
\begin{minipage}{.5\hsize}
\begin{center}
\includegraphics[scale=.5]{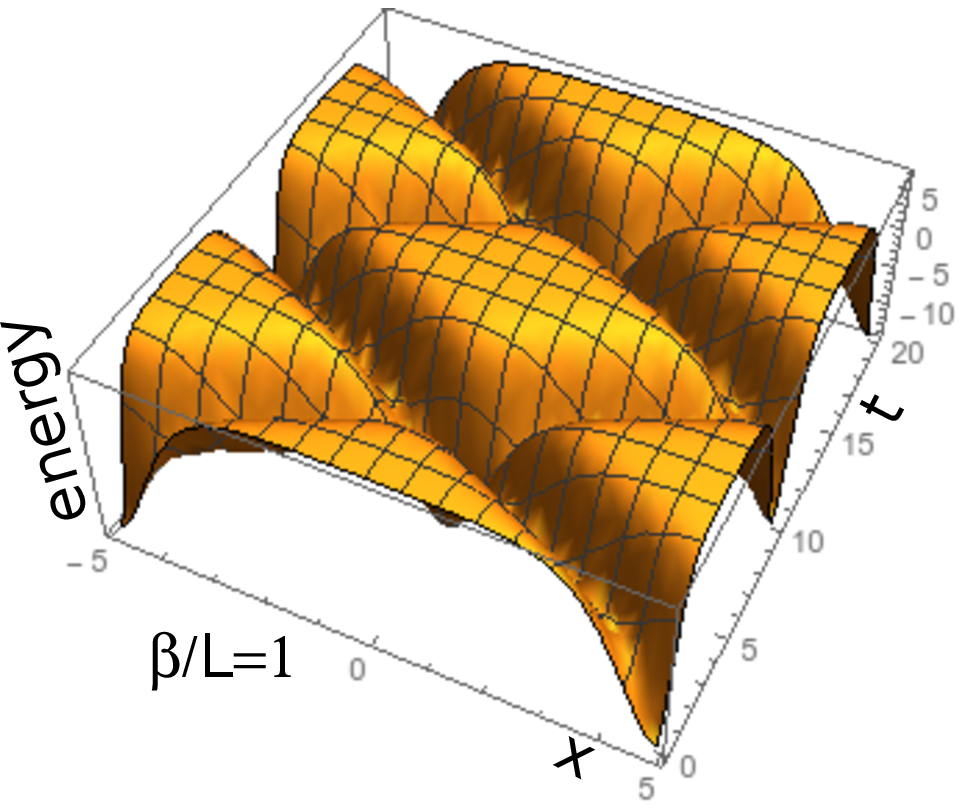}
\end{center}
\end{minipage}
\begin{minipage}{.5\hsize}
\begin{center}
\includegraphics[width=2in, height=1.5in]{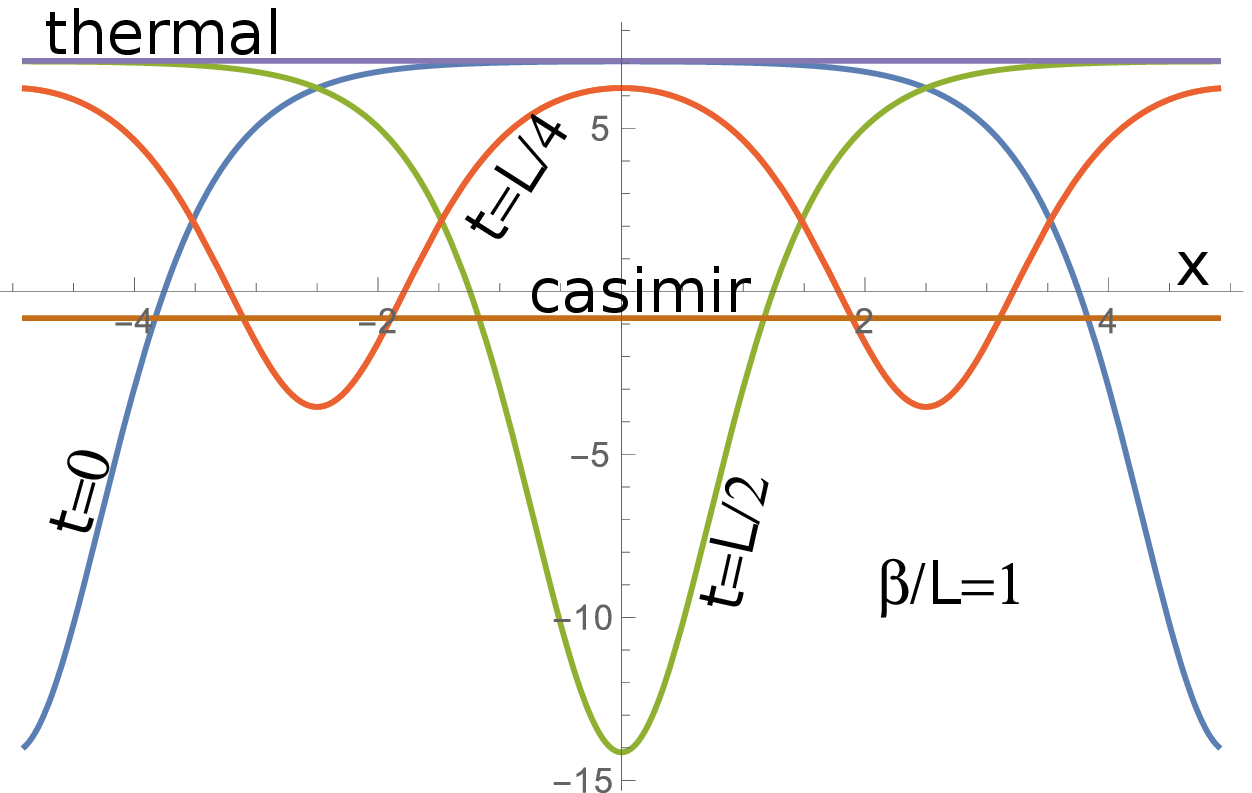}
\end{center}
\end{minipage}
\caption{\footnotesize The time evolution of the normalized energy
  density $\widetilde{\cal E}$ (Eq. \eq{normal-energy}) in the
  quench geometry. In both the figures we have chosen $\b=L=10$. Left panel: The
  time range is taken to be $t \in [0,2L]$.  Note that there are two
  crests, one moving initially to the right (corresponding to
  $T_{++}(x_+)$ of \eq{e-lorentz}), and the other moving to the left
  (corresponding to $T_{--}(x_-)$).  Both are reflected at the wall at $t=
  (n+1/2)L, n \in \mathbb{Z}$. Right panel: Here we display, in a 2D
  plot, some of the features of the 3D plot at $t=0, L/4, L/2$. At
  $t=0$ there is a single local thermal region in the middle. As time
  progresses, it splits into two separate regions (see the curve for
  $t=L/4$), reaching the two ends at $t=L/2$. After this time, the two
  regions turn back and merge at $t=L$, as is clear from the 3D plot on
  the left.}
\label{fig:energy-t}
\end{figure}

\subsubsection{\label{sec:plasma-ball}Space dependence: 
Localized thermal region}

In Figure \ref{fig:energy-2D}, we plot the energy density at a fixed
time $t= 0$ as a function of $x$ for various values of the ratio
$\b/L$, with temperature increasing from left to right. 
In the figure, we see that for temperatures $T\ \gsim\ 1/L$,
the energy density profile has a localized region in the middle where
it agrees with the thermal density. In Section \ref{sec:bulk},
we will provide a holographic interpretation of this observation.

\begin{figure}[H]
\begin{minipage}{0.3\hsize}
\begin{center}
   \includegraphics[width=1.8in,height=1.5in]{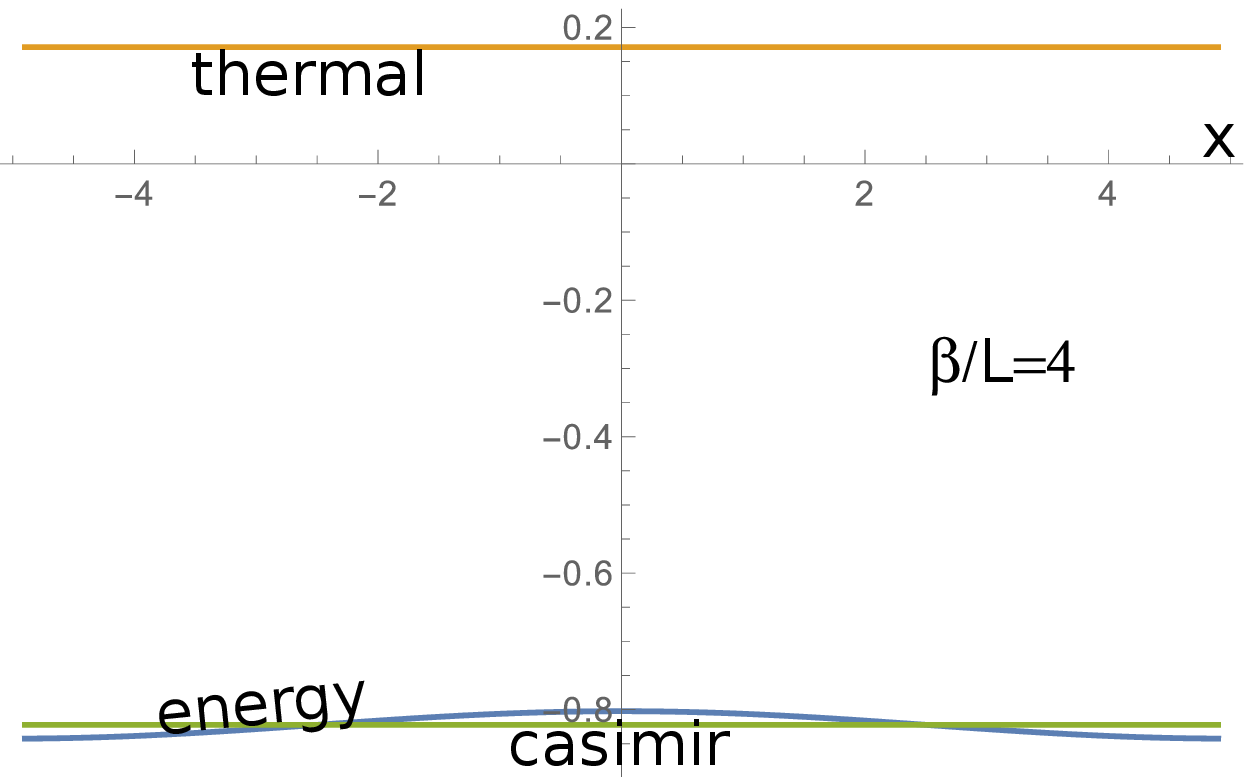}
\end{center}
\end{minipage}
\begin{minipage}{0.3\hsize}
\begin{center}
\includegraphics[width=1.8in,height=1.5in]{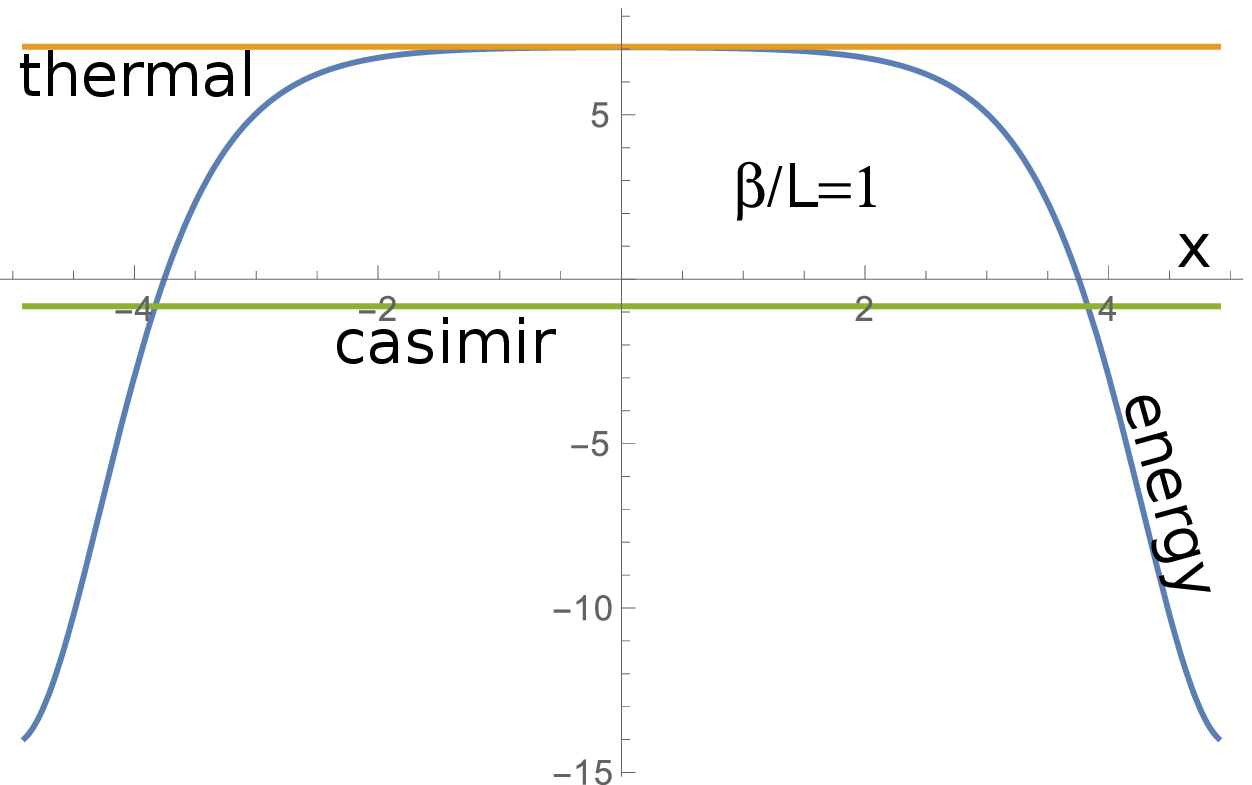}
\end{center}
\end{minipage}
\begin{minipage}{0.3\hsize}
\begin{center}
   \includegraphics[width=1.8in,height=1.5in]{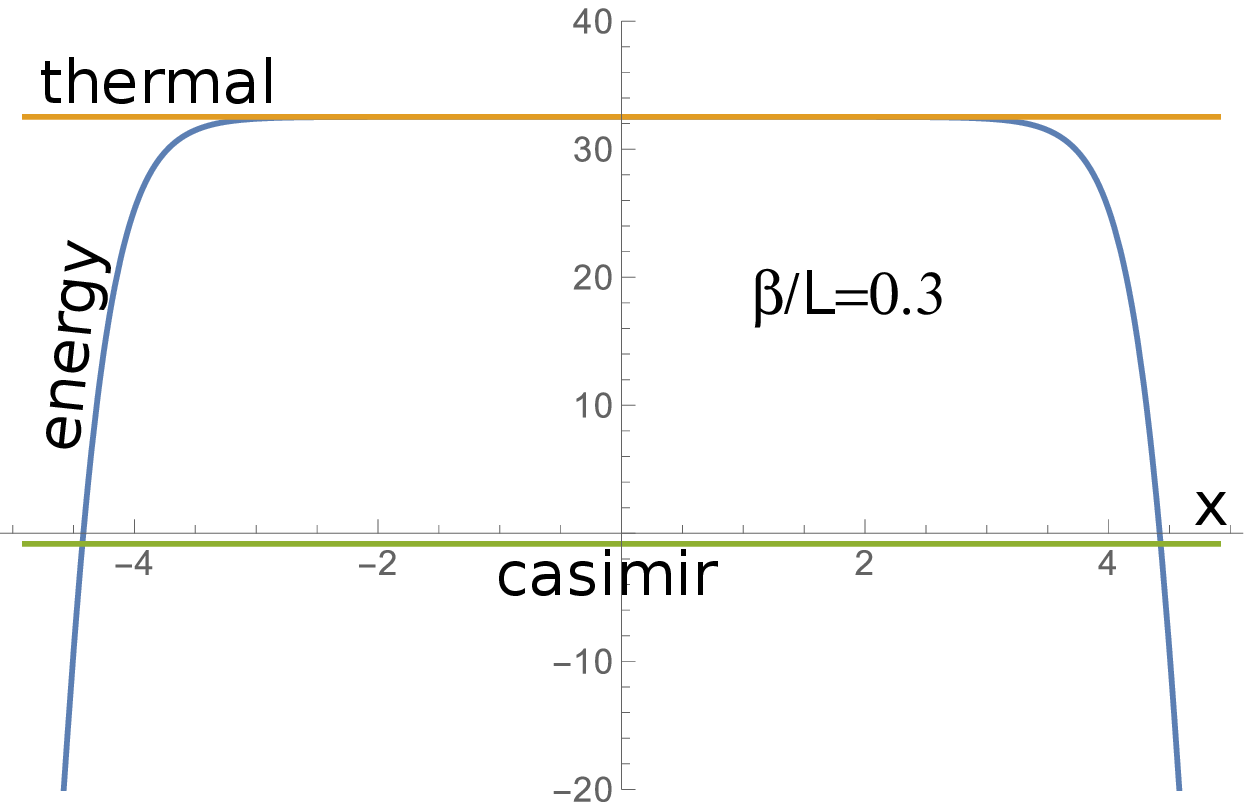}
\end{center}
\end{minipage}
\caption{\footnotesize Plot of the normalized energy density
  $\widetilde{\cal E}(x,t)$ at $t=0$ for various values of
  $\f{\beta}{L}=4,2,0.3$. Left: $\f{\beta}{L}= 4$. This can be
  interpreted as low temperature or small `box size' $L$. The energy
  density approaches the uniform limiting value, the Casimir energy
  density computed in \eq{casimir} in the low temperature limit
  $\f\b{L}=\infty$. Middle: $\f{\beta}{L}= 2$. The energy density
  approaches the thermal value in a small region near the
  middle. Right: $\f{\beta}{L}= \f{1}{4}$. This can be interpreted as
  a high temperature or large box size $L$. The energy density matches
  in a large region the uniform thermal energy density computed in
  \eq{thermal} in the limit $\f\b{L}=0$. In the first panel, the
  one-point functions actually have a periodicity with a period $2L$
  (see Section \ref{sec:period-EE}).}
\label{fig:energy-2D}
\end{figure}

\subsection{\label{sec:one-pt-x-t} Thermalization and `revival' of a 
local operator}

In this section, we discuss some features of time and
position dependence of one-point functions, using the formulae derived
in Section \ref{sec:one-pt}. We consider primary operators of the form
$O_{h,h}(x,t)$ whose one-point functions are given by
\eq{connected-2pt-primary} with $\tau=it$. We also consider a
particular quasiprimary operator $T\bar T(x,t)$, whose one-point
function is given by \eq{TTbar}. The results are
presented in Figure \ref{fig:one-pt}. An important feature of the
time-dependence of these operators, for a given $x$ (see the left
panel of Figure \ref{fig:one-pt}), is that there is a time window,
$t_1< t <t_2$, in which all these local operators `thermalize',
i.e., they approach their `thermal' values:
\begin{align}
\lan O(x,t) \ran \to   \lan O(x) \ran_\b + a \exp[- \g (t-t_1)]
\label{decay}
\end{align}
In case of primary operators the thermal average $\lan O(x) \ran_\b$
vanishes, whereas for $T\bar T$, it is non-zero. It can be seen that
the thermalization rate $\g$ is given by
\begin{align}
\g= 2\pi \Delta /\b,\;   \Delta = h+ \bar h = 2h
\label{gamma}
\end{align}
As $t$ exceeds $t_2$ (which is of order $L/2$ for the operator $T\bar
T$), the one-point function starts deviating from the thermal value
and eventually goes back to its original value at $t= n L$ for some
integer $n$ (this is to be compared with the periodic behaviour 
termed as `revival' in \cite{Cardy:2014rqa}).

\begin{figure}[H]
\begin{minipage}{0.31\hsize}
\begin{center}
   \includegraphics[width=2in,height=2.0in]{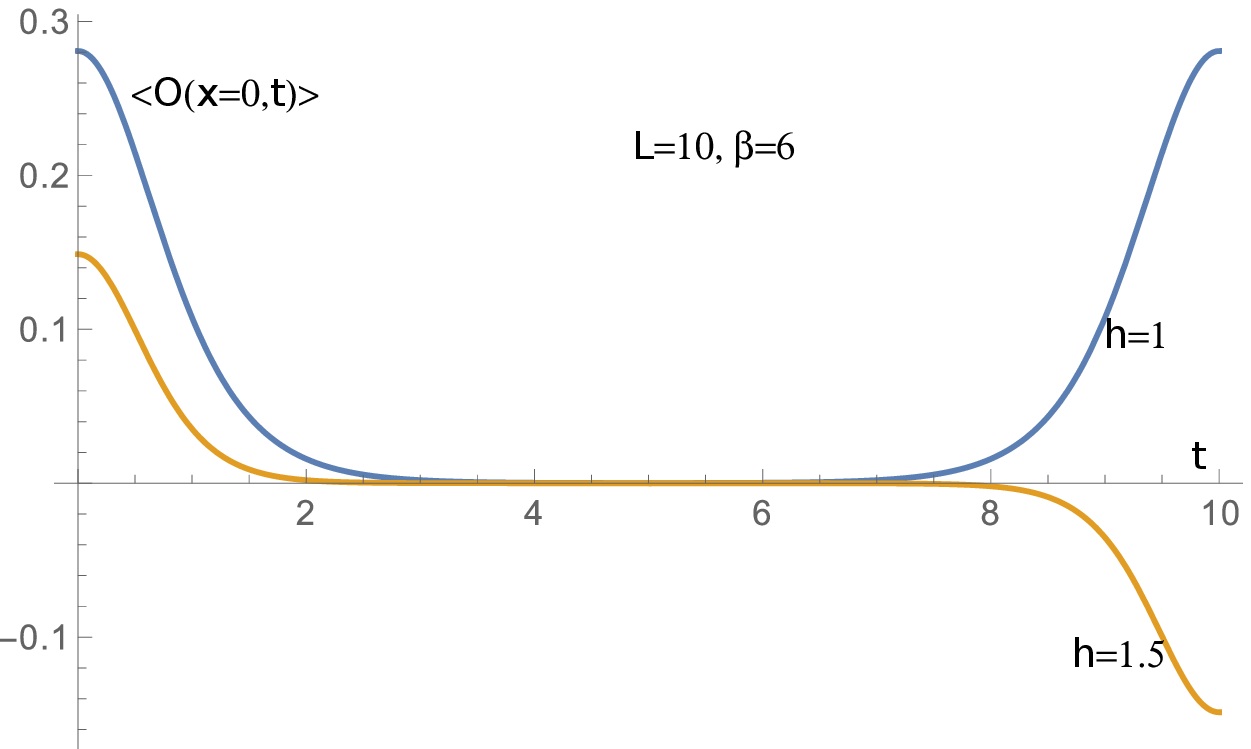}
\end{center}
\end{minipage}
\begin{minipage}{0.31\hsize}
\begin{center}
   \includegraphics[width=1.8in,height=2.0in]{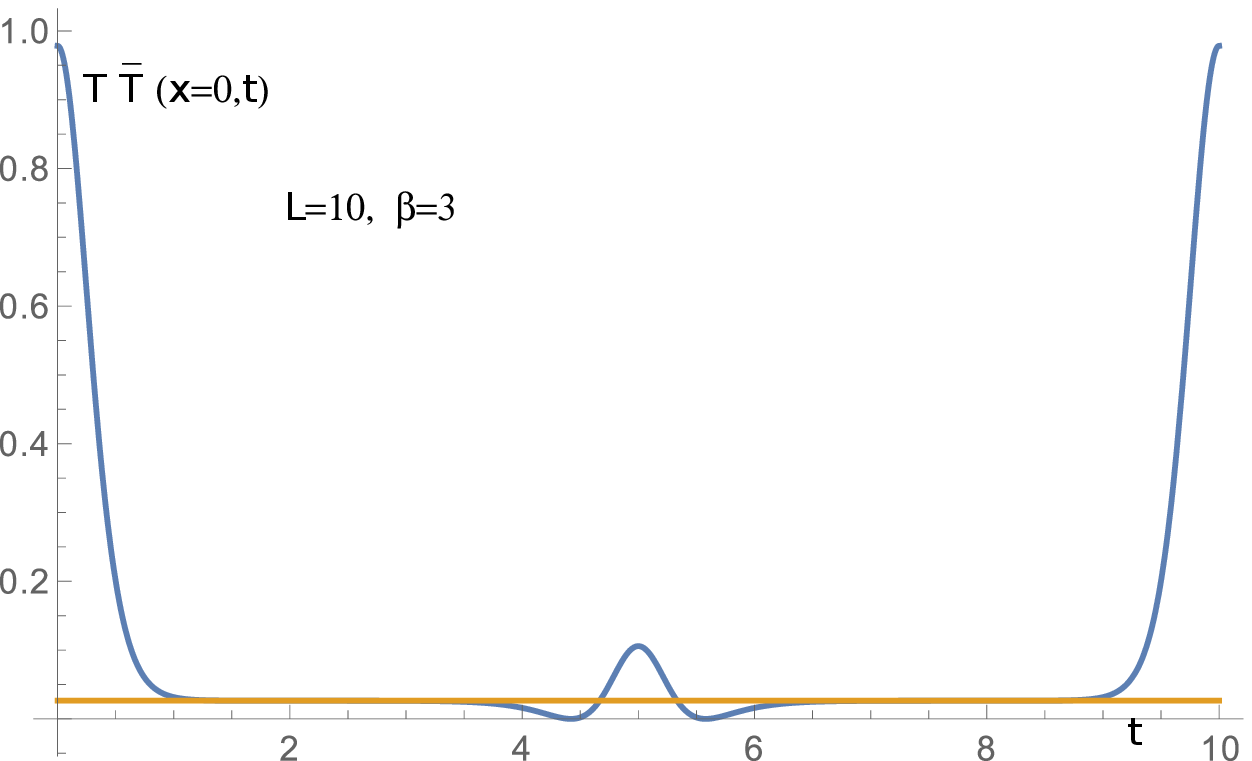}
\end{center}
\end{minipage}
\begin{minipage}{0.31\hsize}
\begin{center}
\includegraphics[width=2in,height=2.0in]{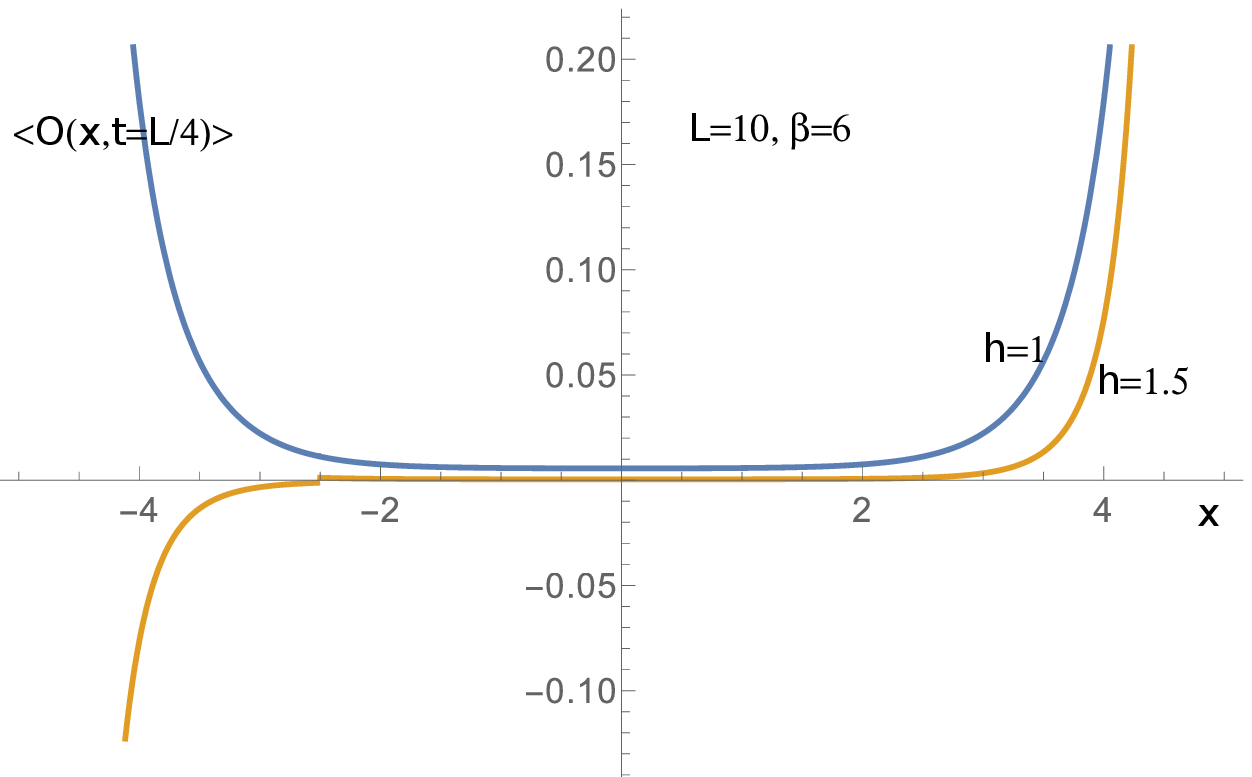}
\end{center}
\end{minipage}
\caption{\footnotesize Plot of one-point functions as a function of
  $x$ and $t$. The left panel shows the time-dependence of one-point
  functions of two primary fields $\lan O(x,t) \ran$ with $h=\bar h=1$
  (blue), and $h=\bar h=3/2$ (orange) at a fixed spatial position
  $x=0$ (see \eq{connected-2pt-primary}). The right panel shows the
  $x$-dependence of the same operators at a fixed $t=L/4$, which
  displays a homogeneous intermediate region. The middle panel shows
  $\lan :T \bar T:(x,t)\ran$ at $x=0$ as a function of $t$ (see
  \eq{TTbar}). The left and middle panels show the exponential decay
  at intermediate times to the thermal value and the eventual revival,
  at $t=L/2$ \cite{Cardy:2014rqa}. In the left panel the thermal value
  is zero. In the middle panel, the horizontal (orange) line
  represents the thermal value, which is non-zero since $:T \bar T:$
  is a quasiprimary operator. More generally, arbitrary one-point
  functions show a time periodicity of $2L$ (see the derivation in
  Section \ref{sec:period-EE}).}
\label{fig:one-pt}
\end{figure}

\section{\label{sec:EE}Evolution of entanglement entropy 
following a quench} 

In this section we would like to compute the time-evolution of
entanglement entropy (EE) of a single interval in a CFT with boundaries, following
a quantum quench (as described in Section \ref{sec:CFT}). We will 
follow up the CFT calculation in this section by a holographic computation
in Section \ref{sec:bulk}, and a computation using a quasiparticle picture
in Section \ref{sec:quasiparticle}.

Let us consider a 2D CFT, defined by the wavefunction \eq{psi-0} 
representing a quantum quench. As explained in Section \ref{sec:CFT},
the norm of the wavefunction \eq{psi-t}, evolved over Euclidean time
$\tau= \b/4$ \footnote{Quantities depending on real time are obtained
  by continuing to $\tau= \b/4 + it$.}, is given by a functional
integral over a Euclidean rectangle $\Sigma$. Let us define the
reduced density matrix for a spatial interval $ A = \{-l/2, l/2\}$ by
\begin{align}
\rho_A = \f{1}{\sqrt{Z}} \;\tr_{A^c} \left|  \psi_0 \ran \lan  \psi_0 
\right|,  \quad \tr (\rho_{A})=1
\label{rho-A}
\end{align}
where $Z$ is a normalization factor. The EE for this interval is then defined as 
\be
S_{A} = - \tr \rho_A \ln\rho_A
\ee
A standard procedure to compute this quantity, called the
replica trick, is to first calculate the Renyi entropy
\begin{align}
S_A^{(n)} \equiv  \f1{1-n} \tr (\rho_A^n)
\label{renyi}
\end{align}
and then take the limit (after analytically continuing $n$ to real values).
\begin{align}
S_{A} = \lim_{n\to 1} S_A^{(n)} =  -\f{\p}{\p n}\big|_{n=1}   \tr \rho_A^n
\label{replica}
\end{align}
Computing \eq{renyi} involves evaluating the partition function on
the manifold $\Sigma_n$, which is an $n$-fold cover of the rectangle
$\Sigma$ branched over $A$. As shown in \cite{Calabrese:2004eu}, this
amounts to computing a two-point function, on the rectangle, of the so
called nth order twist and anti-twist operators, $\sigma_+(w,\bar w)$ and $\sigma_-(w,\bar w)$ respectively, 
each with a conformal dimension $h_{n} =\bar h_n= \frac12 \Delta_n= \f{c}{24}(n-\f{1}{n})$. Thus,
\begin{align}
\tr \rho_A^n = \langle \sigma_+(w_1, \bar w_1)\: 
\sigma_-(w_2, \bar w_2)\rangle_{\text{rect}}
\label{twist}.
\end{align}
where $(w_i, \bar w_i), i=1,2$ are the complex coordinates
\eq{analytic} of the
end-points of the interval $A$. Let us choose the position of the end-points of the interval at a Lorentzian time $t$ to be, 
\begin{align}
w_1=-\f{l}{2} -t,\hspace{.5cm}\bar w_1=-\f{l}{2} +t,\hspace{.5cm} w_2=\f{l}{2} -t,\hspace{.5cm}
\bar w_2= \f{l}{2} +t
\label{w-s}
\end{align}
This 2-point function is obtained by pulling back a corresponding
2-point function on the upper half plane by the conformal maps
(\ref{inverse-map1}), (\ref{map2}), in a generalization of Section
\ref{sec:one-pt}. Hence,
\be
\tr \rho_A^n=\prod_{i=1}^{2}\bigg(\f{dz_i}{dw_i}\bigg)^{h_n}
\bigg(\f{d\bar z_i}{d\bar w_i}\bigg)^{\bar{h}_n}
\langle\sigma_+(z_1,\bar z_1)\sigma_-(z_2,\bar z_2)\rangle_{\text{UHP}}
\ee

To evaluate such a 2-point function on the UHP one needs to use the
method of images.  However, there are subtleties associated with this
method which limits its range of validity. We shall point them out in
the next section before we proceed to calculate the above 2 point
function using the method.

\subsection{Method of Images}

In \cite{Cardy:1984bb}, Cardy showed that an $n$-point function on the
UHP satisfies the same differential equation (corresponding to a Ward
Identity) as a $2n$-point function of purely holomorphic operators on
the full complex plane $\mathbb{C}$.  The two-point function $\langle
\sigma_+(z_{1},\bar z_1) \sigma_-(z_2,\bar z_2) \rangle_{\rm UHP}$ is
thus {\it related}, in this sense, to the 4-point function
\begin{align}
\langle \sigma_+(z_{1})\sigma_-(\bar z_1) \sigma_-(z_2)
\sigma_+(\bar z_2)\rangle_C=
\left( \f{(z_{1}-\bar{z}_{2})(z_{2}-\bar{z}_{1})}{(z_{1}-z_{2})
(\bar{z}_{1}-\bar{z}_{2})(z_{1}-\bar{z}_{1})(z_{2}-\bar{z}_{2})}\right)^{
\Delta_{n}} F(\eta), 
\label{four-pt}
\end{align} 
where function $F(\eta)$ is a (non-universal) function of the cross
ratio
\begin{align}
\eta= \frac{(z_{1}-\bar{z}_{1})(z_{2}-\bar{z}_{2})}{(z_{1}-
\bar{z}_{2})(z_{2}-\bar{z}_{1})}
\label{cross}
\end{align}
The relation mentioned above does not imply, however, that the
two-point function on the UHP is always {\it equal} to the 4-point
function on the plane, since, for one thing, the UHP correlator
involves information about the boundary condition on the boundary of
the UHP whereas the planar correlator does not involve any such
information\footnote{We thank Justin David and Tadashi Takayanagi
for crucial discussions on this issue.}. 

Having said this, it turns out that there are certain limits where the
two correlators are, in fact, essentially \footnote{In this
  subsection, we use the word ``essentially'' to mean up to an
  irrelevant proportionality constant.} equal. One of them is the high
temperature limit where both the 2 and the 4-point functions factorize
in one way or the other. In such a case, the UHP correlator does turn
out to be essentially independent of the boundary state. For
intermediate temperatures, however, there is no such factorization and
the two-point function on the UHP is different from the 4-point
function on $\mathbb{C}$ due to the presence of the boundary.

There is another limit, as we will study, in which the equality holds
(in fact, at all temperatures this time). This is the large central
charge limit.  As we shall see in Section \ref{large-c}, in this
limit, the four point function is easily evaluated and becomes
essentially equal to the two-point function on the UHP. Here too
the information about the boundary condition is lost in the
large $c$ limit.


\subsection{Large $L/\b$ limit}

In the limit of large system size $L$\footnote{Defined in the sense of the limit (a) of footnote \ref{ftnt:limits}.}, the conformal map is given by
(\ref{high-limit}). The computation of EE described above reduces, in
this limit, to the analysis of \cite{Calabrese:2005in,
  Hartman:2013qma}.  The complex coordinates \eq{w-s} are mapped to
\begin{align}
z_1= i e^{-\frac{2\pi}{\beta}(-\f{l}{2}-t)}, \; 
\bar z_1= -i e^{-\frac{2\pi}{\beta}(-\f{l}{2}+t)}, \;
z_2=i e^{-\frac{2\pi}{\beta}(\f{l}{2}-t)},\; \bar z_2= 
-i e^{-\frac{2\pi}{\beta}(\f{l}{2}+t)}
\label{z-high}
\end{align}
on the upper half plane. Here we have assumed that 
$t/l \ll L$. Therefore, the 
cross ratio $\eta$ in \eq{cross} becomes
\begin{align}
\eta= \f{2\cosh^2 \f{2\pi t}\b}{\cosh \f{2\pi l}\b +\cosh\f{4\pi t}\b}
\label{cross-high}
\end{align}
When $t/\b,l/\b\gg 1$, 
the cross ratio behaves as
\be
\eta\simeq \f{e^{\f{4\pi t}{\beta}}}{e^{\f{4\pi t}{\beta}}+e^{\f{2\pi l}{\beta}}}
\ee
It is easy to verify that,
\begin{align}
\eta \xrightarrow{} \left\{ \begin{array}{l} 0 ~ \hbox{for}~t < l/2 \\ 
 1 ~ \hbox{for}~t > l/2 \end{array}
\right.
\label{high-value}
\end{align}
where we have assumed $|t-l/2| \gg \b$.
Both the limits of $\eta$ correspond to a factorization of the four-point
function \eq{four-pt} into a product of two-point functions in the
two crossed channels, respectively. This leads to a simpler calculation of
the EE which gives 
\begin{align}
S_{A}= \left\{ \begin{array}{l} \f{2c\pi t}{3 \beta} ~ \hbox{for}~t< l/2 \\ 
\f{c\pi l}{3 \beta}  ~ \hbox{for}~t>l/2 \end{array}
\right.
\label{EE-high}
\end{align}
where a divergent constant has been subtracted \cite{Calabrese:2005in}.  

These results are universal since they depend only on the central
charge of the CFT.  The behaviour \eq{EE-high} represents the linear
rise of the EE followed by the saturation to the thermal value (see
Figure \ref{fig:surface1}, left part of the third panel). By the
periodicity properties mentioned in Section \ref{sec:period-EE},
we can also show that CFT results reproduce the right part of
the third panel of Figure \ref{fig:surface1}. One way
to see this is to note that for  $L-t, l \ll L$ the map \eq{inv-map2}
reduces to a map similar to \eq{z-high}, with $t \to L-t$. 



\subsection{\label{sec:CFT-low}Low temperature ($L/\b$ small)}

In the low temperature limit\footnote{Defined in the sense of limit
  (c) of footnote \ref{ftnt:limits}.}, the conformal map is given by
(\ref{low-limit}). Using this, the $w$-coordinates \eq{w-s} are mapped
to
\begin{align}
z_1 &= i \cot\bigg(\f{\pi}4 + \f{\pi(-l/2 - t)}{2L}\bigg), \;
\bar z_1= -i \cot\bigg(\f{\pi}4 + \f{\pi(-l/2 + t)}{2L}\bigg)\;
\nonumber\\
z_2 &= i  \cot\bigg(\f{\pi}4 + \f{\pi(l/2 - t)}{2L}\bigg), \hspace{.5cm}
\bar z_2= -i  \cot\bigg(\f{\pi}4 + \f{\pi(l/2+ t)}{2L}\bigg).
\label{z-low}
\end{align}
The cross-ratio \eq{cross} now becomes 
\begin{align}
\eta= \cos^2\f{\pi l}{2 L}
\label{cross-low}
\end{align}
Thus, unlike in the high temperature limit, the cross ratio does not take
a special value. Depending on the size of the interval $l$, it can lie anywhere 
between $0$ and $1$. Consequently, the four-point function \eq{four-pt} does
not factorize into two-point functions. This implies that
the CFT EE remains a non-universal quantity, depending on the particular CFT under  consideration. 

\paragraph{A puzzle} The above remark immediately raises the following
puzzle. We will see in Section \ref{sec:hEE}, the holographic EE,
as computed using \eq{ryu} and \eq{ryu-new}, is independent of the
details of the dual CFT  (except for the dependence on the central charge 
$c$ which determines the Newton's constant through \eq{newton}).
\footnote{\label{ftnt:bh}This is a recurrent
theme in AdS/CFT. A similar observation is: many different CFT states 
appear to evolve into
states described by black holes. The emergence of universality in the
bulk in that context is related to the universality of thermal physics.} 

The puzzle is that while the holographic EE is universal, the CFT EE
certainly does not appear to be so. How does one reconcile this with
AdS/CFT?
Also, what is the CFT calculation which will agree with the hEE
at all temperatures and exhibit universality?

\paragraph{Resolution}
This puzzle will be resolved in the next subsection,
by appealing to the \emph{large central charge} limit.

\subsection{\label{large-c}Universality of 
the Entanglement Entropy at large central charge}
\subsubsection{At Low Temperature}
One way of obtaining universal results for the EE, (as will be shown
in the low temperature bulk calculations), from the CFT, is by looking
at the large $c$ systematics (similar issues have been addressed
in \cite{Leichenauer:2015xra}). As has been mentioned, the calculation of
$\tr \rho_A^n$ on the manifold $\Sigma_n$, can be mapped to the
calculation of the two point correlator of n-th order twist and
anti-twist operators on the UHP.  Using the method of images, this is
equivalent to a four point function of two twist and two anti-twist
operators on $\mathbb{C}$.
The four point function is then,
\begin{align}
\tr\rho_A^n=\prod_{i=1}^{2}\left(\f{dz_i}{dw_i}\right)^{h_n}
\!\!\left(\f{d\bar z_i}{d\bar w_i}\right)^{\bar h_n}\!\! {\cal G}_4,\;
\;\; {\cal G}_4 \equiv 
\lan\sigma_+(z_1)\sigma_-(\bar z_1)\sigma_+(\bar z_2)
\sigma_-(z_2)\ran_{\mathbb{C}}
\nonumber
\end{align}
For convenience, we send the four points $(z_1,\bar z_1,z_2,\bar z_2)$ to
$(\infty,1,0,\eta)$, such that
\be
{\cal G}_4=(z_{1\bar 2}z_{\bar 1 2})^{-2 h_n}G_n(\eta),\hspace{2cm}
G_n(\eta)=\lan\sigma_+(\infty)\sigma_-(1)\sigma_+(\eta)\sigma_-(0)\ran_{\mathbb{C}}
\label{4-point}
\ee
Consider the scaled 4 point function in the $\eta\rightarrow0$ channel. In the large central charge limit, all the 
conformal blocks exponentiate (\cite{Zamolodchikov-87}). The form of the function is,
\be
G_n(\eta)=\sum_p a_p e^{-\f{nc}{6}f_p(h_n,\eta;nc)}
\ee
Recent results (\cite{Hartman:2013mia}) suggest that it is only the vacuum block that dominates the above sum over intermediate 
operators. In that case, the 4-point function is,
\be
G_n(\eta)  \approx e^{-\f{nc}{6}f_0(h_n,\eta;nc)}
\ee
Taking the four points $z_{1,2}, \bar z_{1,2}$ to be given by 
\eq{z-low}, we get the cross-ratio to be,
\be
\eta=\f{z_{1\bar 1}z_{2 \bar 2}}{z_{1\bar 2}z_{2\bar 1}}=\cos^{2}\f{\pi l}{2 L}
\label{cross-ratio}
\ee
From (\cite{Hartman:2013mia}), one knows that the result for the vacuum block (in the $\eta\rightarrow0$ channel) is,
\be
f_0(h_n,\eta;nc)=12 \alpha \log (\eta)+\mathcal{O}(\alpha^2)
\ee
where $\alpha=\f1{12}(n-1)(=h_n/c)$. 
Hartman's result tells us that we can extrapolate the result found around $\eta=0$ up till $\eta=\f12$ owing to monotonicity of the conformal blocks as a function of $\eta$
for light operator exchanges. A similar statement holds in the other channel around $\eta=1$. 
The vacuum block in that case is,
\be
f_0(h_n,\eta;nc)=12 \alpha \log (1-\eta)+\mathcal{O}(\alpha^2)
\ee
Here, again, the results can again be extrapolated up to $\eta=\f12$. This enables us to state that the 
purely holomorphic Euclidean 4-point function is in fact equal to the 2-point function on the UHP.
Again, the interesting thing to notice is that the information about the boundary gets completely lost in this equivalence at large $c$!

The Entanglement Entropy in the s-channel ($\eta\rightarrow0$) is then,
\be
S_A=\lim_{n\rightarrow1}\f1{1-n}\log Tr \rho_A^n=\lim_{n\rightarrow1}\bigg[\f{4h_n}{1-n}\log(\f{\pi}{2L})-\f{n c}{6}\f1{1-n}12 \alpha\log(\eta)\bigg] 
\ee
where the first term comes from the Jacobian and the prefactor in \ref{4-point}. With $\alpha=h_n/c$ and $\eta=\cos^2(\f{\pi l}{2 L})$, we get,
\be
S_A=\f{c}3\log \bigg(\f{2L}{\pi a}\cos{\f{\pi l}{2L}}\bigg)
\ee
This result is, of course, valid up to $\eta=1/2$ and hence, for $L/2\leq l<L$. 
In the other channel $\eta\rightarrow1$,
\be
S_A=\f{c}3\log \bigg(\f{2L}{\pi a}\sin{\f{\pi l}{2L}}\bigg)
\ee
This is valid for $0\leq l\leq L/2$.
In both the above cases, we have introduced the UV cut-off $a$ by hand to regulate the answers. As, we will see later, these answers match exactly with the 
holographic results in the low-temperature limit.

\subsubsection{For All Temperatures}

The results obtained at low temperatures for large c are much more general. In fact, we would like to show the equivalence of these results with the lengths of the bulk geodesics,
which can be calculated for all temperatures. The analysis of the four point function on $\mathbb{C}$ is completely general. The information about the initial nature of the geometry 
(a rectangle in this case), is contained in the pull-back maps and hence in the Jacobian factors. Only the cross-ratio on the plane knows about the initial geometry through the 
pull back maps. Let us calculate the entanglement entropy on the plane and use an unspecified conformal transformation to pull it back onto a non-trivial geometry. 

On the plane,
\be
\tr\rho_A^n=\prod_{i=1}^{2}\left(\f{dz_i}{dw_i}\right)^{h_n}
\!\!\left(\f{d\bar z_i}{d\bar w_i}\right)^{\bar h_n}\!\! {\cal G}_4\;
\ee
Let us call the Jacobian factors $J=\prod_{i=1}^2(\f{dz_i}{dw_i})(\f{d\bar{z}_i}{d\bar{w}_i})$ for simplicity. Then (as we saw in the previous section),
for the $\eta\rightarrow1$ channel,
\begin{eqnarray}
S_A&=&\lim_{n\rightarrow1}\f1{1-n}\bigg[h_n\log\bigg(\f{a^4 J}{(z_{1\bar 2}z_{\bar 1 2})^2}\bigg)-\f{nc}6 12\alpha\log\eta)\bigg]\nonumber\\
&=&\lim_{n\rightarrow1}\f{2h_n}{n-1}\bigg[\log\bigg(\f{z_{1\bar 2}z_{\bar 1 2}}{a^2\sqrt{J}}\bigg)+n\log(\eta)\bigg]\nonumber\\
&=&\f{c}6\log\bigg(\f{\eta}{a^2}\f{z_{1\bar 2}z_{\bar 1 2}}{\sqrt{J}}\bigg)
\label{large-c-EE}
\end{eqnarray}
Note, the factor of $a^4$ in the first line of the expression has been introduced for dimensional reasons and has a length dimension.
With the identifications, $c=\f3{2 G_3}$ and $\eta=\f{z_{1\bar 1}z_{2 \bar 2}}{z_{1\bar 2}z_{2\bar 1}}$, the above EE is,
\be
S_A=\f1{4G_3}\log\bigg(\f{z_{1\bar 1}z_{\bar 2 2}}{a^2\sqrt{J}}\bigg)
\ee
This expression for the entanglement entropy exactly coincides the sum of the bulk geodesic lengths between the 
boundary points $(z_1,\bar z_1)$ and $(z_2,\bar z_2)$ (with the identifications $z_{\pm}=f(x_{\pm})$), in the geometry corresponding to the pulled back (non-trivial) surface, at all temperatures.
This is the `disconnected' channel in \eq{ryu-new}.

Similarly, when $\eta\rightarrow0$, we shall have to use the conformal block in the other channel. Now, $(1-\eta)=\f{z_{1 2}z_{\bar 1 \bar 2}}{z_{1\bar 2}z_{\bar 1 2}}$. Using this,
the entanglement entropy is,
\begin{eqnarray}
S_A&=&\lim_{n\rightarrow1}\f1{1-n}\bigg[h_n\log\bigg(\f{a^4 J}{(z_{1\bar 2}z_{\bar 1 2})^2}\bigg)-\f{nc}6 12\alpha\log(1-\eta)\bigg]\nonumber\\
&=&\f{c}6\log\bigg(\f{(1-\eta)}{a^2}\f{z_{1\bar 2}z_{\bar 1 2}}{\sqrt{J}}\bigg)\nonumber\\
&=&\f1{4G_3}\log\bigg(\f{z_{12}z_{\bar 1 \bar 2}}{a^2\sqrt{J}}\bigg)
\end{eqnarray}
This matches with the bulk entropy in the `connected' channel in \eq{ryu-new} with geodesics joining the points $(z_1,z_2)$
and $(\bar{z}_1,\bar{z}_2)$.
\vspace{0.3 cm}

\section{\label{sec:bulk}Bulk dual}

In this section we discuss a holographic dual to the above two
dimensional quenches \cite{Ugajin:2013xxa}.  It is known that a class
of two dimensional CFTs with a large central charge $c$ have an
equivalent description in terms of gravity in  three dimensional
anti de Sitter space (AdS$_{3}$). In Poincar\'e \ coordinates the
metric of AdS$_{3}$ is given by
\begin{align} 
ds^{2}=\f{d\zeta^2+dz_+ dz_-}{\zeta^2}. 
\label{poincare}
\end{align}
Here $\zeta=0$ is the conformal boundary of $AdS_{3}$. The boundary
coordinates $z_\pm= z_1 \mp z_0$ describe the plane where the dual
(Lorentzian) CFT lives.  We will also consider the Euclidean
continuation of the above metric where we will set $z_\pm= \mp i\{z,
\bar z\}$ (see \eq{analytic-z}, Appendix \ref{zplus-minus}). The corresponding 
dual CFT on the complex $z$-plane will then be Euclidean.

A holographic description of the CFT on a space with boundaries (BCFT)
has been proposed in \cite{Karch:2000ct, DeWolfe:2001pq,
  Takayanagi:2011zk, Fujita:2011fp, 
Nozaki:2012qd}. According to this
proposal, the holographic dual of the BCFT on the upper half plane
$\{(z_1,z_0)| z_0 >0 \}$ is given by the $z_0 >0$ region of the
Poincare $AdS_{3}$. The $z_0=0$ plane serves as a boundary of the
holographic spacetime\footnote{\label{ftnt:brane-tmunu}In a recent 
proposal \cite{Magan:2014dwa} a stress-tensor has been ascribed to this plane
as in the case of a D-brane.}, with the boundary condition that the extrinsic
curvature of the boundary vanishes $K_{\mu \nu}=0$\footnote{This
  boundary condition has a one parameter generalization
  \cite{Takayanagi:2011zk,Fujita:2011fp} which corresponds to
  available boundary conditions in the BCFT.}, and ends (at $\zeta=0$)
on the boundary of the UHP. A schematic picture is presented in Figure
\ref{fig:BCFT}.

\begin{figure}[H]
\begin{center}
\includegraphics[width=3.5in, height=2.5in]{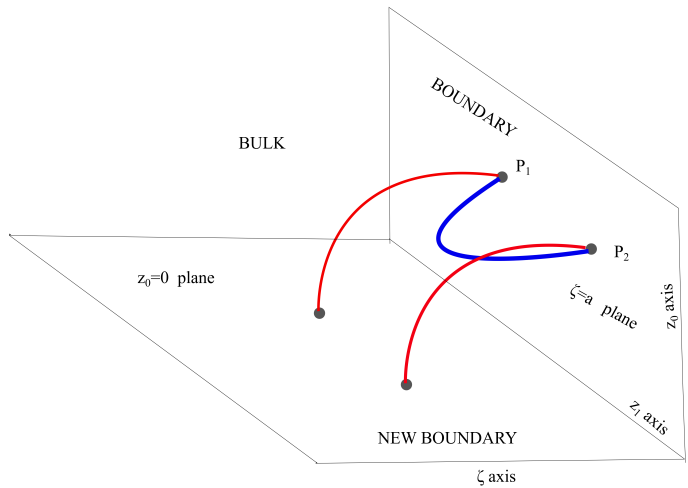}
\end{center}
\caption{\footnotesize Holographic dual to BCFT on the UHP (upper half
  plane).  A one-point function in the UHP is given by an extremal
  geodesic emanating from the relevant boundary point and ending on
  the boundary plane (the length of this is half of the geodesic
  connecting the boundary point and its image). Entanglement entropy
  of the interval connecting two boundary points $P_1$ and $P_2$ is
  given by the minimum \eq{minimum} of the length of extremal
  geodesics connecting these points and their images pairwise. We call
  the `blue' and 'red' configurations the `connected' and
  `disconnected' channels respectively.}
\label{fig:BCFT}
\end{figure}

The above description gives the bulk dual of a Lorentzian CFT on the
UHP; we would, however, like to obtain the bulk dual for a CFT on a
manifold \eq{lorentz-man}, $M_L=$ interval $\times \mathbb{R}$.  Now,
in Appendix \ref{zplus-minus} we will find a map from the manifold $M_L$
to the UHP, given by \eq{fp-fm-app}:
\begin{align}
z_\pm =  f_\pm(x_\pm),\hspace{.5cm} f_+(x) = f_-(x)= f(x) \equiv   
- b\ \f{sn\left[\f{4K(b^4)}{\beta}(x-L/2), 1-b^4\right]}{cn
\left[\f{4K(b^4)}{\beta}(x-L/2),1- b^4\right]} 
\label{lorentz}
\end{align}

\paragraph{Bulk dual through large diffeomorphisms}

Our strategy to find a bulk dual to the CFT on $M_L$ will be to find a
solution to Einstein's equations with a negative cosmological constant
whose boundary is $M_L$. This problem can be solved by the method
introduced in \cite{Roberts:2012aq}, where it was shown (building on
the works of \cite{Brown:1986nw} and \cite{Banados:1998gg}) how
solutions diffeomorphic to the AdS$_3$ geometry \eq{poincare} (which
are hence solutions to Einstein equations) can be found where the
diffeomorphism is non-trivial at the boundary and reduce to the map
\eq{lorentz}.  The diffeomorphisms alluded to here are analogous to
`large gauge transformations' \cite{Regge:1974zd, Wadia:1979yu} and
are called `large' diffeomorphisms or `solution generating
diffeomorphisms' \cite{Mandal:2014wfa}. Of course, large
diffeomorphisms which reduce to a specific conformal transformation
$f_\pm$ are not unique since one can always compose with any local
diffeomorphism which reduces to identity at the boundary.  This
ambiguity can be removed, however, if one demands that the resulting
bulk metric is in the Fefferman-Graham gauge. These ideas have been
used in the study of quantum quenches in \cite{Ugajin:2013xxa,
  Mandal:2014wfa,Roberts:2012aq}.  The quantitative form of `large'
diffeomorphism was found by (\cite{Roberts:2012aq}) and is given by
\begin{align}
\zeta=4 z \f{\left(f'_+ f'_-\right)^{3/2}}D, \; z_\pm
=  f_\pm(x_\pm) - \f{2 z^2 {f_\pm'}^2 f_\mp''}D,\; D= 4 f'_+f'_-
+  z^2 f''_+ f''_- 
\label{roberts}
\end{align}
Note, as a check, that as $\zeta \to 0$ (the boundary of the Poincare
metric \eq{poincare}), we have $z \to 0$ (assuming that $D$ remains
finite), and  $z_\pm \to  f_\pm(x_\pm)$. It is important to note
that for functions $f_\pm(x_\pm)$ which are not one-to-one and
have vanishing Jacobians at the boundaries of fundamental domains,
one must restrict the above transformation \eq{roberts} to a
given choice of fundamental domain. 

Applying this large diffeomorphism to the Poincar\'{e} metric, we get the following metric
\cite{Banados:1998gg}
\begin{align}
ds^2= \f{dz^2}{z^2}+  \f14\left(L_{+}dx_{+}^2+L_{-}dx_{-}^2\right) 
+\left(\f{1}{z^2}+
\f{z^2}{16}L_{+}L_{-}\right)dx_{+}dx_{-}. 
\label{banados}
\end{align}
where $L_{\pm}$ are given in terms of Schwarzian derivatives
\eq{def-sch} of the boundary conformal map
\begin{align}
L_{\pm}=-2 \{z_\pm, x_\pm\}=
\f{3f''^2_{\pm}-2f'_{\pm} f_{\pm}'''}{f'^2_{\pm}}
\label{lp-lm}
\end{align}
The metric (\eq{banados}) is called the Fefferman-Graham metric and
has been written in the Fefferman Graham gauge. Note that as $z\to 0$,
the leading terms of this metric reduce to that of the Poincar\'e
metric \eq{poincare}.  The two metrics, however, represent physically
different solutions of the theory due to the existence of the
subleading terms of the metric \eq{banados} involving the $L_\pm$
terms which represent nontrivial stress tensors at the boundary
\cite{Balasubramanian:1999re}. These $L\pm$'s capture the `surface
charges' of \cite{Brown:1986nw} characterizing the quantum state. The
precise relation (\cite{Banados:1998gg},\cite{Balasubramanian:1999re})
between $L_\pm$ and the CFT stress tensors is,
\begin{align}
T_{\pm,\pm}(x_\pm)= \f{L_\pm}{16 G_3}= \f{c}{24}  L_\pm
\label{t-vs-l}
\end{align}
where we have used the following relation between the central charge
of the CFT and the 3D Newton's constant 
\begin{align} 
c= \f{3}{2G_{3}}.
\label{newton}
\end{align}

\paragraph{Properties of the metric:}

The metric \eq{banados}, therefore, provides the promised solution to
the Einstein's equations (with $\Lambda<0$) whose conformal boundary
coincides with the manifold $M_L$. Hence, gravity on this metric
provides the geometric dual to the CFT on $M_L$ (which is the CFT of
our interest).

We have shown in Appendix \ref{app:poincare-btz} the high temperature
limits and low temperature limits of the geometry \eq{banados}
represent the BTZ black hole and the global AdS respectively. We can
show that at sufficiently high temperatures the geometry contains a
horizon. For the purposes of the rest of the paper, the important
property of the metric \eq{roberts} is that the spatial direction $x$
is compactified with a periodicity $L$ (i.e. equal to the
spatial size $L$).  This follows as a consequence of the periodicity
properties of the Elliptic functions (see, e.g. \eq{periods} and
similar statements for the Lorentzian map). 

The periodicity properties of the CFT observables can be
holographically interpreted in terms of the above periodicity property
of the bulk metric. Thus, e.g. a two-point function, which can be
specified by the geometric properties of a geodesic with end-points at
the boundary, are periodic because the geodesic will come back to
itself after a time period $L$. We will find below an explicit
example of how this happens in case of the holographic
EE. 

\paragraph{Other candidate bulk duals}

It is important to note that in related contexts, somewhat different
proposals for bulk dual geometries have appeared. For example, the
bulk dual to the large $L$ limit of the CFT studied here, used by
\cite{Hartman:2013qma} is obtained by first dividing the Penrose
diagram of an eternal BTZ black string geometry vertically by an
end-of-the world brane and then taking the top half of it as the
relevant bulk dual geometry. We are currently investigating the
relation between our proposal and this geometry. The hEE computed in
both geometries turn out to be the same. Some other proposals for bulk
geometries dual to quantum quench involve the AdS-Vaidya metric,
see, e.g. \cite{Balasubramanian:2011ur}.

\subsection{Holographic quantum quench and entanglement entropy  
\label{sec:hEE}}

As mentioned in Section \ref{sec:bulk}, for a class of 2D CFTs with a
large central charge $c$, a holographic description is available in
terms of a weakly coupled gravity dual in asymptotically AdS$_3$
spaces. A computation of the CFT partition function over the branched
cover $\Sigma_n$ maps to a computation of the bulk partition function
over a dual geometry whose conformal boundary coincides with
$\Sigma_n$ \cite{Hartman:2013mia,Faulkner:2013yia,Lewkowycz:2013nqa}.
As shown in \cite{Lewkowycz:2013nqa}, this observation leads to the
well-known Ryu-Takayanagi formula \cite{Ryu:2006bv,Hubeny:2007xt} for
the holographic entanglement entropy ({\bf hEE}) of an interval $A$
\begin{align}
S_{hol,A}= {\rm{ext}}~ \f{l(\gamma_{A})}{4G_{3}},
\label{ryu}
\end{align}
where $\g_A$ is the extremal curve in the bulk ending at the boundary
of the interval $A$, $l(\g_A)$ is the length of this curve, and
$G_{3}$ is Newton's constant which is related to the central charge
$c$ of the dual CFT by \eq{newton}.
The extremum is taken among all curves $\gamma_{A}$ which are
homotopic to the subsystem A. 

Let us first compute \eq{ryu} in the bulk dual of the BCFT on the
upper half plane. The dual geometry is the upper half of the
spacetime \eq{poincare}. Suppose that the boundary of the subsystem A
consists of two points $P_1=z_{\mu,1}$$=(z_{+,1},
z_{-,1})$,$P_2=z_{\mu,2}$$=(z_{+,2}, z_{-,2})$, $\mu=+,-$. In this
case we have two extremal {\it geodesics}, as shown in Figure
\ref{fig:BCFT}:
\\ (i) One is a geodesic connecting $P_1$ and $P_2$, which 
we call the connected geodesic, whose length is 
\begin{align} 
l(\g_A)_{c}=2\log \f{|P_1-P_2|}{\zeta_{min}} =2\log \f{\sqrt{(z_{+,1}-z_{+,2})
(z_{-,1}-z_{-,2})}}{\zeta_{min}}, \label{eq:con}
\end{align}
where $\zeta_{min}$ denotes the UV cut-off in the CFT which,
by the rules of AdS/CFT, corresponds
to placing the asymptotic boundary at $\zeta=\zeta_{min}$ (this 
regulates the extremal surface area which will diverge
otherwise). $| P_{1}-P_{2}|$ denotes the
distance measured in the flat boundary metric $ds^2= dz_+ dz_-$.  
\\
(ii) The presence of the additional spacetime boundary at $z_0=0$,
leads to the existence of  
an extra geodesic consisting of two independent 
geodesic segments, each of which connects $P_1$ (or $P_2$) to this 
new boundary.
We will call this the disconnected geodesic; the length of
each segment, say the first one, is half of that of
a geodesic connecting $P_1$ to each its `image point' $P'_1$
below the boundary (similarly with $P_2$). 
Combining the two segments, we get 
\begin{align}
l(\g_A)_{dc}=\log \f{|P_1-P'_1|}{\zeta_{min}}  
+  \log \f{|P_2-P'_2|}{\zeta_{min}} 
\label{eq:discon}
\end{align}
The entanglement entropy \eq{ryu} is determined by taking the minimum
among these \cite{Hubeny:2007xt}.
\begin{align}
S_{hol,A} = \f1{4 G_3}\times  
{\rm min} \{l(\g(A))_{c}, l(\g_A)_{dc} \}
\equiv  {\rm min} \{ S_{c},  S_{dc} \}
\label{minimum}
\end{align}

\paragraph{Fluctuation of the UV cut-off with a large diffeomorphism}

The construction here gives the hEE for the bulk dual of a CFT on the
UHP, which is  \eq{poincare} with an additional boundary. As
we found in Section \ref{sec:bulk}, the bulk dual to the BCFT on the
rectangle, which is of our original interest, is given by the geometry
\eq{banados}. Since the latter metric is related to the former
by a diffeomorphism (which is non-trivial at the boundary), the
extremal geodesics in the latter metric  can be
obtained by pulling back (\ref{eq:con}), (\ref{eq:discon}) into the
geometry.  This prescription is known to reproduce the time evolution
of entanglement entropy in various quantum quenches
\cite{Sotiriadis:2008, Ugajin:2013xxa}. The effect of the
large diffeomorphism can be represented by a fluctuation of the
end-points $P_1, P_2$ \cite{Ugajin:2013xxa, Mandal:2014wfa}, while the
inside geometry remains identical. The effect of this is that
the point $P_1$, represented by the
coordinates $(\zeta_{min},z_{\pm,1})$ in the Poincare geometry
\eq{poincare} is transformed
to  $(z_{min},x_{\pm,1})$ according to \eq{roberts}. We define the
original CFT to be that on the rectangle, with a lattice cut-off $a$.
In AdS/CFT, this instructs to introduce a UV cut-off surface 
$z_{min}=a$ in the geometry \eq{banados}.
Near the horizon, using the $z\to 0$ limit of \eq{roberts}, we have
\begin{align}
\zeta_{1, min}= a \sqrt{f'_+(x_{+,1}) f'_-(x_{-,1})}
\label{fluct-cutoff}
\end{align} 
and similarly for the point $P_2$. In other words, insisting on
a given cut-off in the original CFT leads to a local definition
of the UV cut-off, as above. The prescription for generalization of formulae
like \eq{eq:con} to take this into account is simple
 \cite{Ugajin:2013xxa, Mandal:2014wfa}: just replace $\zeta_{min}$
in \eq{eq:con} by $\sqrt{\zeta_{1,min} \zeta_{2,min}}$.

\paragraph{Expression for the hEE}

Using these ingredients we get the following  formulae
for the extremal length of  two geodesics in the new bulk geometry
\eq{banados}:
\begin{align}
l(\g_A)_{c} & = 
\log \f{(f(x_{+,1}) -  f(x_{+,2}))(f(x_{-,1}) -  f(x_{-,2}))
}{a^2 \sqrt{f'(x_{+,1})f'(x_{+,2})\,f'(x_{-,1})f'(x_{-,2})}},
\nonumber\\
l(\g_A)_{dc} &
= \log \f{(f(x_{+,1})-f(x_{-,1}))(f(x_{+,2})-f(x_{-,2}))}{a^2 
\sqrt{f'(x_{+,1})
f'(x_{-,1})\,f'(x_{+,2})
f'(x_{-,2})}} 
\label{ryu-new}
\end{align}  
Here $f_+ = f_-=f$ is as defined in \eq{lorentz}.

The boundary points of interest in this problem are given by
the coordinates (which are Lorentzian versions of the points \eq{w-s})
\begin{align}
x_{\pm,1}= -l/2 \mp t, \; x_{\pm,2}= l/2 \mp t
\label{x-pm}
\end{align}
These represent the end-points of the entangling interval (region
$A$) at time $t$.

\subsection{\label{sec:hEE-t}Evolution of holographic entanglement entropy}

In this section, we compute time evolution of holographic entanglement
entropy in the global quench with boundaries, using the prescription
we mentioned in the previous subsection. We mainly focus on the low
temperature limit $\f{L}{\beta} \rightarrow 0$ as well as the high
temperature limit $\f{L}{\beta} \rightarrow \infty$.  We also compare
the result to the naive CFT entanglement entropy derived by neglecting
the function $F(z)$ in (\ref{four-pt}) which is the theory
dependent part of the four point function (\ref{four-pt}).  We find
that whereas they do not agree in the low temperature limit, they
do agree in the high temperature limit. This suggests that the
behavior of the entanglement entropy, as computed from the bulk dual,
is universal, which coincides with the usual universality (coming
from the factorization limit) at high temperatures, but it is
a {\it new universality} at low temperatures (see 
Section \ref{large-c}). 

\begin{figure}[H]
\begin{minipage}{0.3\hsize}
\begin{center}
   \includegraphics[width=45mm, height=35mm]{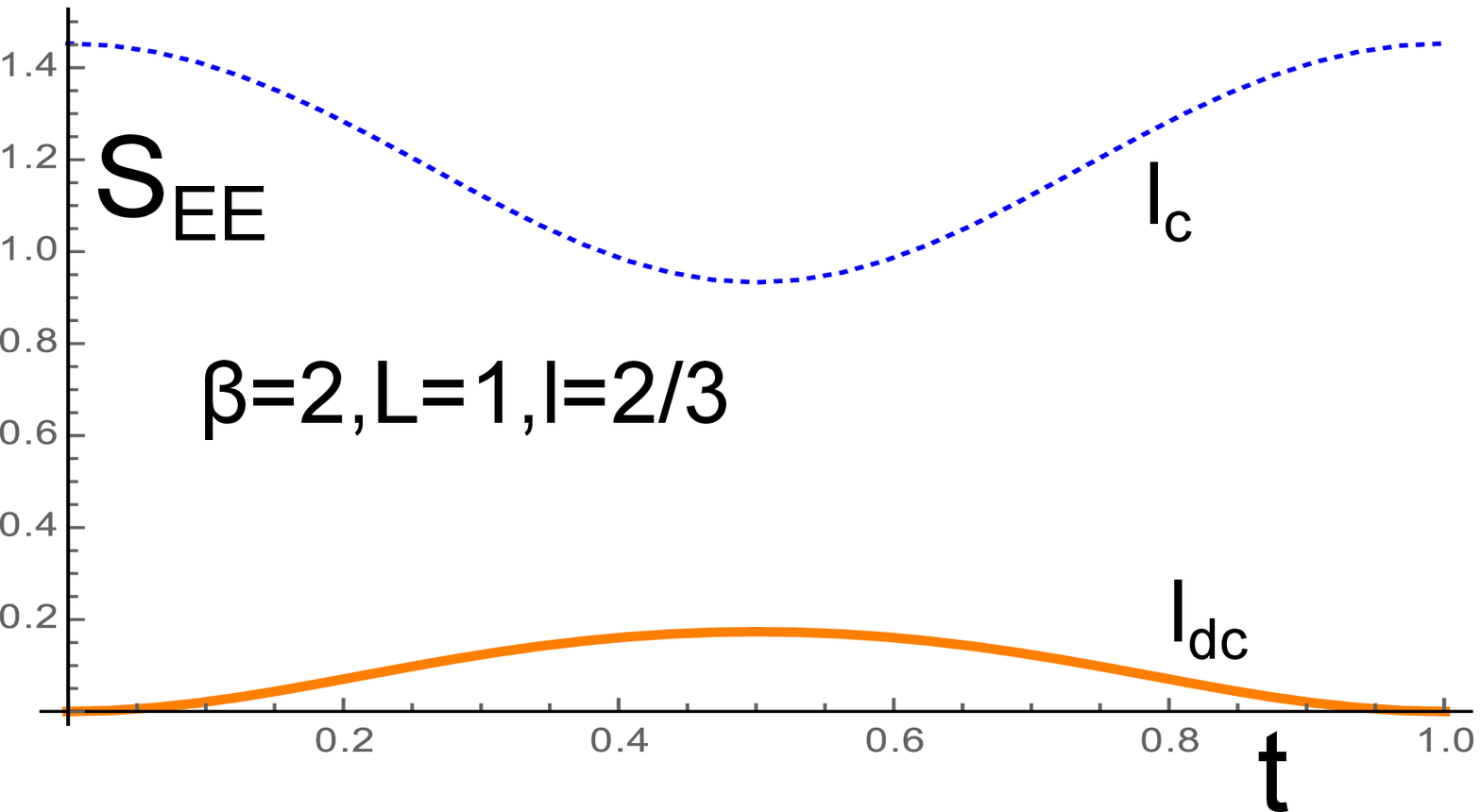}
\end{center}
\end{minipage}
\begin{minipage}{0.3\hsize}
\begin{center}
\includegraphics[width=45mm, height=35mm]{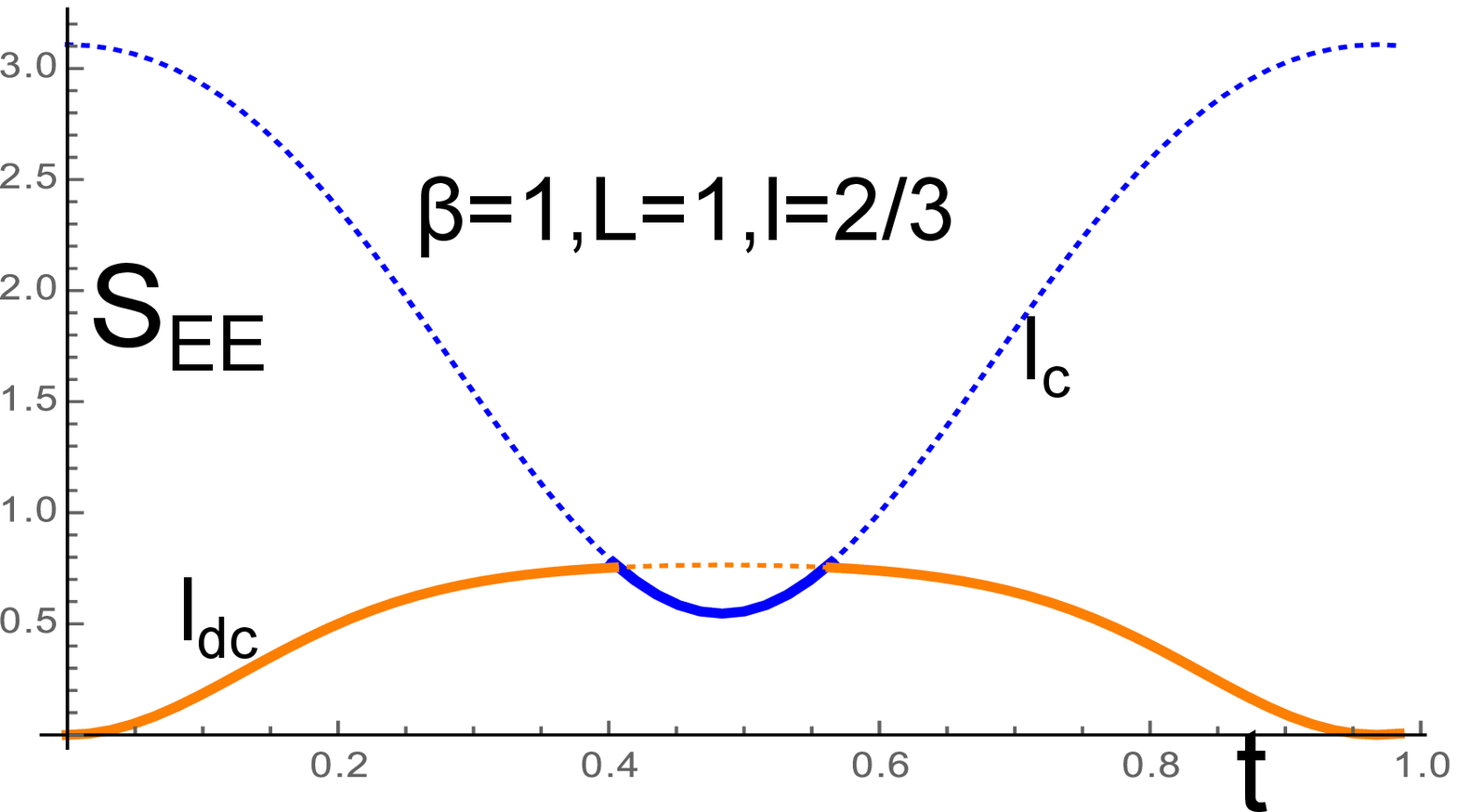}
\end{center}
\end{minipage}
\begin{minipage}{0.3\hsize}
\begin{center}
   \includegraphics[width=45mm, height=35mm]{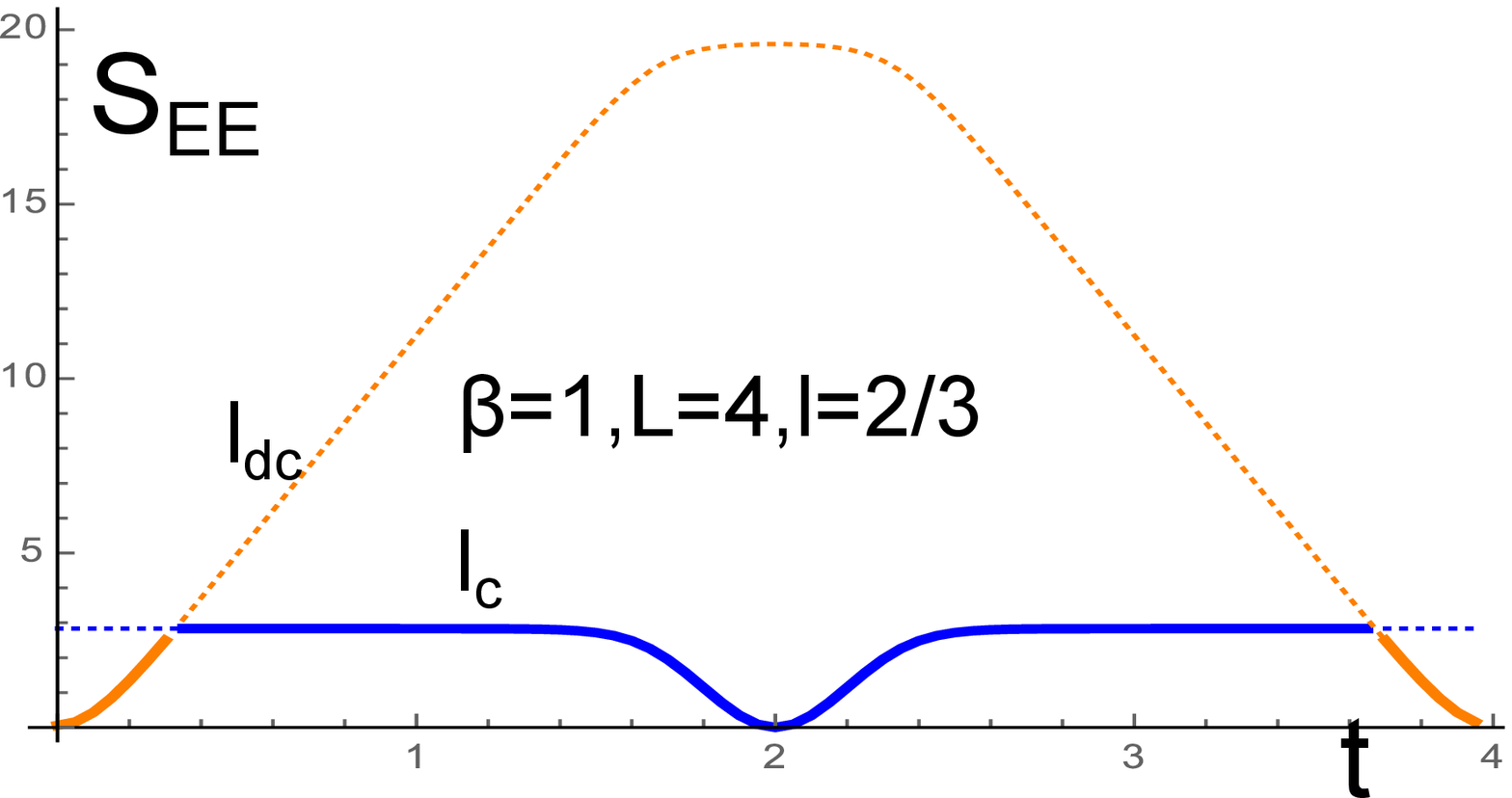}
\end{center}
\end{minipage}
\caption{\footnotesize Plot of the time evolution of the length of
  geodesics which determine, via \eq{ryu}, \eq{minimum}, the
  entanglement entropy ($S_{EE}$) of an interval with end-points at
  $\mp l/2$ (for convenience we have rescaled the geodesic lengths by
  a factor of $\f12$ and adjusted the height of each curve by choosing
  the UV cut-off $a$ in \eq{ryu-new} appropriately). The length of the
  connected path, $l_{c}$ \eq{eq:con} is shown in blue, and that of
  the disconnected, $l_{dc}$ \eq{eq:discon} is shown in red.  The left
  panel corresponds to low temperature with $L/{\beta}=1/6$, the
  middle panel to intermediate temperature with $L/{\beta}=1$ and the
  right panel corresponds to large $L$ (or high
  temperature\protect\footnotemark) with $L/{\beta}=4$. By the minimum
  prescription of \cite{Hubeny:2007xt}, \eq{minimum}, $S_{EE}$ is
  given by the curve segments which lie lower; these are represented
  by solid lines in the figures. The curve segments lying higher are
  not relevant in the computation of EE, and are represented by dotted
  lines. Note the exchange of dominance (`phase transition') for
  intermediate and high temperatures. The above EE is exactly
  reproduced by the CFT calculations in the large $c$ limit (see
  Section \ref{large-c}). For the rightmost panel, except for the
  central dip, the EE can be seen to match the CFT result even without
  the help of large $c$ asymptotics, using factorization into
  appropriate two-point functions at large $L/\b$.}
\label{fig:surface1}
\end{figure}

\footnotetext{Strictly speaking, we distinguish
    between $L \to \infty$ and $\b \to 0$ in the sense of footnote
    \ref{ftnt:limits}.}

\subsubsection{\label{sec:period-EE}Time periodicity of two-point functions and the hEE}

Note that the map $f(x_\pm)$ appearing in the above expressions
satisfies (cf. \eq{periods}) $f(x_\pm)= f(x_\pm + 2L)$.  Since the
spatial locations of the boundary points are fixed as in \eq{x-pm},
the above periodicity implies a periodicity in time $t \equiv t+ 2L$
of two-point functions of operators whose AdS representation is in
terms of geodesics. The periodicity here can be geometrically
understood as follows.  The map \eq{roberts} with periodic $f_\pm=f$
effects a quotienting of the Poincare geometry.  As time progresses,
each end-point of a geodesic climbs along a straight line in the
$x$-$t$ plane; the light-cone coordinates both trace out a periodic
path on the cylinder (which is double the manifold $M_L$
\eq{lorentz-man}), leading to the time-periodicity mentioned above.
  
For the hEE of a centrally located interval (with end-points located
at $x=\mp l/2$), the period, in fact, turns out to be $L$ rather $2L$,
as in Figure \ref{fig:surface1}. This periodicity holds for both the
connected or the disconnected expression ($l_c$ or $l_{dc}$) in
\eq{ryu-new}. To see this, note that the odd (even) parity of the $sn$
(respectively, $cn$) functions imply that $f(x_{\pm,1} +L) = -
f(x_{\mp,2})$, $f(x_{\pm,2} +L) = - f(x_{\mp,1})$. Here we have used the
definitions \eq{x-pm}.  In a similar way we can also show the symmetry
$l_c(t)= l_c(L-t)$, $l_{dc}(t)= l_{dc}(L-t)$, which is evident in
Figure \ref{fig:surface1}.
 
\subsubsection{Low temperature limit  $\f{L}{\beta} \rightarrow 0$}

In this limit the map $f_\pm$ of \eq{lorentz} reduces to the analytic
continuation of \eq{low-limit}, namely
\begin{align}
z_\pm(x_\pm)= \cot\left(\pi/4 + \pi x_\pm/(2 L)\right)
\label{low-limit-an}
\end{align}
By using this conformal map, and the formalism described above,
we can calculate the contribution of the
connected surface to the holographic entanglement entropy $S_{c}$, and
the contribution of the disconnected surface $S_{dc}$.
\begin{align}
S_{c}=\f{c}{3} \log\left[ \f{2L}{a\pi} \sin \left(\f{\pi l}{2L}
  \right) \right], \qquad S_{dc}= \f{c}{3} \log\left[ \f{2L}{a\pi} \cos
  \left(\f{\pi l}{2L} \right) \right]
\end{align}
In this case the entanglement entropy does not depend on time, hence
we are interested in how this depends on the size of the subsystem $A$.
By taking the minimum between $S_{c}$ and $S_{dc}$ we get,
\begin{eqnarray}
S_{hEE}=\left\{ \begin{array}{ll}
 \f{c}{3} \log\left[ \f{2L}{a\pi} \sin \left(\f{\pi l}{2L} \right) \right] 
& l< \f{L}{2}\\
 \f{c}{3} \log\left[ \f{2L}{a\pi} \cos \left(\f{\pi l}{2L} \right) \right] 
& l>\f{L}{2} \\
\end{array} \right.
\end{eqnarray}
When the size of the subsystem is smaller than the half of the
subsystem $l< \frac{L}{2}$, the result is the same as that in the
vacuum state on a cylinder with circumference $2L$.  This is natural
because the local physics is the same in both cases.  When $l>
\frac{L}{2}$, they become different. This is necessary because
entanglement entropy has to vanish if we take the subsystem to be the
total system.

\paragraph{Nontrivial agreement with CFT}

Note that the holographic EE obtained above exactly agrees with the
CFT EE, discussed in Section \ref{large-c} (see \eq{large-c-EE}).  As
explained before, this is a nontrivial agreement, since in general,
including at low temperatures, the CFT EE is, {\it a priori},
non-universal, being given by a four-point function. However, as we
found in Section \ref{large-c}, in an appropriate limit of large
central charge, one recovers a universal result \eq{large-c-EE} 
which depends only on the central charge and no other feature of the
CFT. This then agrees with the hEE obtained above.

\subsubsection{Large $L$ limit  $\f{L}{\beta} \rightarrow \infty$}

If we slightly turn on the temperature, the area of the extremal
surfaces starts to depend on time. In the low temperature regime, the
area of one extremal surface (the disconnected one) is always smaller
than the other one (the connected one) for a fixed subsystem size $l$
(see Figure \ref{fig:surface1}).  However above some value of the
temperature, there are phase transitions.  When $0< t< L/2$, it
happens twice. We can check that for sufficiently large
$\f{L}{\beta}$, the area of disconnected surface is smaller at early
times, but since $S_{dc}$ linearly grows in time, the connected
surface becomes the minimal surface at some critical time. This is the
first phase transition. As we will see in the next section, this
critical time $t_{c}$ depends on the size of the subsystem. When $
l>\frac{L}{2}$, this is given by $t_{c}= \f{L-l}{2}$. When $l<
\frac{L}{2}$, this critical time is $t_{c}=\f{l}{2}$. This is natural
because if we take a small subsystem limit $ \f{l}{L} \ll 1,
\f{L}{\beta} \gg 1$, the result should approach that of the usual
global quenches without boundary walls (see Figure
\ref{fig:surface1}).  As is obvious from the periodicity mentioned in
Section \ref{sec:period-EE}), we encounter a second phase transition
after $t$ crosses $L/2$, when the disconnected surface becomes minimal
again. The central dip in this figure cannot be understood from CFT,
but in the next section on quasiparticles, we have an accurate
understanding of the entire plot of the third panel.

\section{The quasiparticle picture of the 
evolution of the entanglement entropy} \label{sec:quasiparticle}

The evolution of the entanglement entropy at sufficiently 
high temperatures
limit can be interpreted by the free-streaming quasiparticle picture
of \cite{Calabrese:2005in}.
In this picture, one models the quenched state as 
populated by entangled quasiparticles pairs at
every point consisting of a left- and a right-moving particle, 
each moving at the speed of
light. The entanglement entropy of an interval
$A$ increases by $ \f{\pi c}{6 \beta}$
when a quasiparticle goes outside (inside) the interval while its
entangled partner is still inside (outside) the interval.

\subsection{The quasiparticle interpretation for global quench 
for infinite spatial size} \label{sec:qgq}

In this subsection we will 
review the interpretation of the time evolution of the
entanglement entropy in the global quench on the non-compact line
$\mathbb{R}$.  The contribution of the quasi-particle pairs, which are
located in $[-(|x|+d|x|),-|x|]$ initially, to the
entanglement entropy at a fixed time, depends on the value of
$|x|$. If the quasiparticles are located inside the
interval, $A=\{|x|< \f{l}{2}\}$ at the initial time, the contribution to the
entanglement entropy $s^{(1)}_{A}(|x|,t)$ is given by
\be
s^{(1)}_{A}(|x|,t)=\f{c \pi}{6 \beta}\left[\theta \left(t-(-|x|+\f{l}{2}) \right)-\theta \left(t-(|x|+\f{l}{2}) \right) \right]
\ee 
This is because, when $t=-|x|+\f{l}{2}$, the left moving partner of the
entangled pair goes outside the interval, and when $t=|x|+\f{l}{2}$
the right moving partner goes outside the interval.

Similarly, the contribution of the quasi-particles outside the
interval $A$ is
\be
s^{(2)}_{A}(|x|,t)= \f{c \pi}{6 \beta}\left[\theta \left(t-(|x|-\f{l}{2}) \right)-\theta \left(t-(|x|+\f{l}{2}) \right) \right]
\ee
The total  entanglement entropy $S_{A}(l,t)$ for the interval 
$A$ is then given by 
\be
S_{A}(l,t)=  2\int^{\infty}_{\f{l}{2}}s^{(2)}_{A}(|x|,t) d|x| +2\int^{\f{l}{2}}_{0} s^{(1)}_{A}(|x|,t) d|x|.
\ee
Here we incorporate the factor 2 to include the contribution of $x>0$ region.
Performing the integral we get
\be
S_{A}(l,t)=\f{2c\pi t}{3\beta} \theta \left(\f{l}{2}-t \right)+\f{c\pi l}{3\beta} \theta \left(t-\f{l}{2} \right). \label{eq:glo}
\ee
which gives us the result from a CFT calculation \cite{Calabrese:2004eu,
  Calabrese:2005in}.

\subsection{The evolution of entanglement entropy in the presence of 
boundaries}

We would now like to discuss the time evolution of the entanglement
entropy in the presence of the boundaries. The new physics in this case
comes from the reflection of the quasiparticles
off the boundary walls.The
entanglement entropy consists of two contributions. One is the
contribution from the inside of the interval, and the other is the
contribution from the outside of the interval.  As we will see below
the form of the each contribution is further classified by the ratio
$\f{l}{L}$ of size of the interval $l$ to the size of the total
system $L$.

Let us consider the motion of the left moving and the right moving
quasiparticles which are located inside the interval $A$, say $-|x|
\in A$ at the initial time $t=0$. The left mover goes outside $A$
at $t=\f{l}{2}-|x|$, then it bounces off the boundary wall
at $t=\f{L}{2}-|x|$, and becomes a right mover. When
$t=L-|x|-\f{l}{2}$, it enters the interval $A$ again. Similarly the
right mover goes outside the interval $A$ at $t=\f{l}{2}+|x|$, bounces
off the boundary wall at $\f{L}{2}+|x|$, and enters the interval
again at $t=L+|x|-\f{l}{2}$.

\subsection{ $\f{l}{L}>\f{1}{2}$}

Let us consider the contribution of the quasiparticle pairs which are
located in $[-(|x|+d|x|),-|x|]$ at the initial time $t=0$ to the
entanglement entropy when $\f{l}{L}>\f{1}{2}$. The contribution of 
those
inside the interval depends on the precise value of $|x|$.  For $|x|<
\f{L-l}{2}$, since $|x|+ \f{l}{2} <(L-|x|-\f{l}{2})$,
\begin{align}
s^{{\rm in}(1)}_{A}&=\f{c \pi}{6 \beta}\left[\theta\left( t-(\f{l}{2}-|x|) \right)-\theta\left( t-(\f{l}{2}+|x|) \right)\right] \nonumber \\ 
&+\f{c \pi}{6 \beta}\left[\theta\left( t-(L-|x|-\f{l}{2}) \right)-\theta\left( t-(L+|x|-\f{l}{2}) \right)\right]
\end{align}
When $\f{L-l}{2}< |x| < \f{l}{2}$, since $ (L-|x|-\f{l}{2})< |x|+\f{l}{2}$
\begin{align}
s^{{\rm in}(2)}_{A}&=\f{c \pi}{6 \beta}\left[\theta\left( t-(\f{l}{2}-|x|) \right)-\theta\left( t -(L-|x|-\f{l}{2})\right)\right] \nonumber \\ 
&+\f{c \pi}{6 \beta}\left[\theta\left( t -(\f{l}{2}+|x|)\right)-\theta\left( t-(L+|x|-\f{l}{2}) \right)\right]
\end{align}
The contribution of the quasiparticles located outside the
interval is
\begin{align}
s^{{\rm out}}_{A}&=\f{c \pi}{6 \beta}\left[\theta\left( t-(|x|-\f{l}{2}) \right)-\theta\left( t -(L-|x|-\f{l}{2})\right)\right] \nonumber \\ 
&+\f{c \pi}{6 \beta}\left[\theta\left( t -(\f{l}{2}+|x|)\right)-\theta\left( t-(L+|x|-\f{l}{2}) \right)\right]
\end{align}
The total entanglement entropy is then given by 
integrating all contribution of the quasiparticles. 
\be
S_{A}(l,t)=2\int^{\f{L-l}{2}}_{0}s^{{\rm in}(1)}_{A} d|x|+ \int^{\f{l}{2}}_{\f{L-l}{2}} S^{{\rm in}(2)}_{A}(|x|,t)+\int^{\f{L}{2}}_{\f{l}{2}} s^{{\rm out}}_{A}(|x|,t) d|x|
\ee
The form of the integral of $s^{{\rm in}(2)}_{A}$ depends on whether
$\f{l}{L }>\f{3}{4}$ or not. The sum of the remaining two terms 
also depends on the size of the interval, ie $\f{l}{L}>\f{3}{4}, \quad
\f{3}{4}>\f{l}{L}>\f{2}{3} $ or $\f{2}{3}>\f{l}{L}>\f{1}{2}$.
At the end of the day, these different cases give the same net
entanglement entropy.

\begin{displaymath}
S_{A}(l,t)=\f{\pi c}{6 \beta} \times \left\{
\begin{array}{l}
2t,\quad (0<t<\f{L-l}{2}) \quad L-l, \quad (\f{L-l}{2}<t<\f{l}{2}) \quad -2t+L, \quad (\f{l}{2}<t<\f{L}{2}) \\
2t-L \quad (\f{L}{2}<t<\f{2L+l}{2}) \quad L-l, (\f{2L+l}{2}<t<\f{2L-l}{2}), \quad -2t+2L, (\f{L+l}{2}<t<L)
\end{array}
\right.
\end{displaymath}

In figure \ref{fig:ee38}, we plot the EE obtained above from the
quasiparticle picture and compare it to the holographic result.  It is
clear that the comparison works rather well.

It is important to note that in this case $l>\f{L}{2}$, and the
entanglement entropy is not thermalized at any time, i.e. it does not
become proportional to the interval size $l$.

\subsection{ $\f{l}{L}<\f{1}{2}$}

The contribution from inside the interval is now given by
\begin{align}
s^{{\rm in}}_{A}&=\f{c \pi}{6 \beta}\left[\theta\left( t-(\f{l}{2}-|x|) \right)-\theta\left( t-(\f{l}{2}+|x|) \right)\right] \nonumber \\ 
&+\f{c \pi}{6 \beta}\left[\theta\left( t-(L-|x|-\f{l}{2}) \right)-\theta\left( t-(L+|x|-\f{l}{2}) \right)\right]
\end{align}
The contribution of the region   $\f{l}{2}<|x|<\f{L-l}{2}$ is
\begin{align}
s^{{\rm out}(1)}_{A}&=\f{c \pi}{6 \beta}\left[\theta\left( t-(|x|-\f{l}{2}) \right)-\theta\left( t -(|x|+\f{l}{2})\right)\right] \nonumber \\ 
&+\f{c \pi}{6 \beta}\left[\theta\left( t -(L-|x|+\f{l}{2})\right)-\theta\left( t-(L+|x|-\f{l}{2}) \right)\right]
\end{align}
The contribution of the region is $\f{L-l}{2}<|x|< \f{L}{2}$ is 
\begin{align}
s^{{\rm out}(1)}_{A}&=\f{c \pi}{6 \beta}\left[\theta\left( t-(|x|-\f{l}{2}) \right)-\theta\left( t -(|x|+\f{l}{2})\right)\right] \nonumber \\ 
&+\f{c \pi}{6 \beta}\left[\theta\left( t -(L-|x|+\f{l}{2})\right)-\theta\left( t-(L+|x|-\f{l}{2}) \right)\right]
\end{align}

\begin{figure}[H]
\begin{minipage}{0.5\hsize}
\begin{center}
\includegraphics[width=50mm]{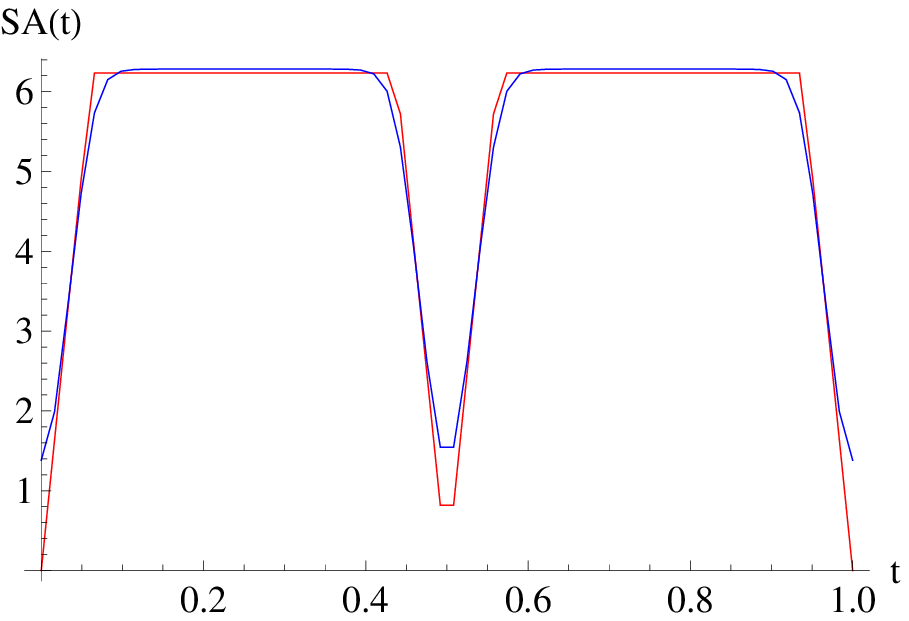}
\end{center}
\end{minipage}
\begin{minipage}{0.5\hsize}
\begin{center}
\includegraphics[width=50mm]{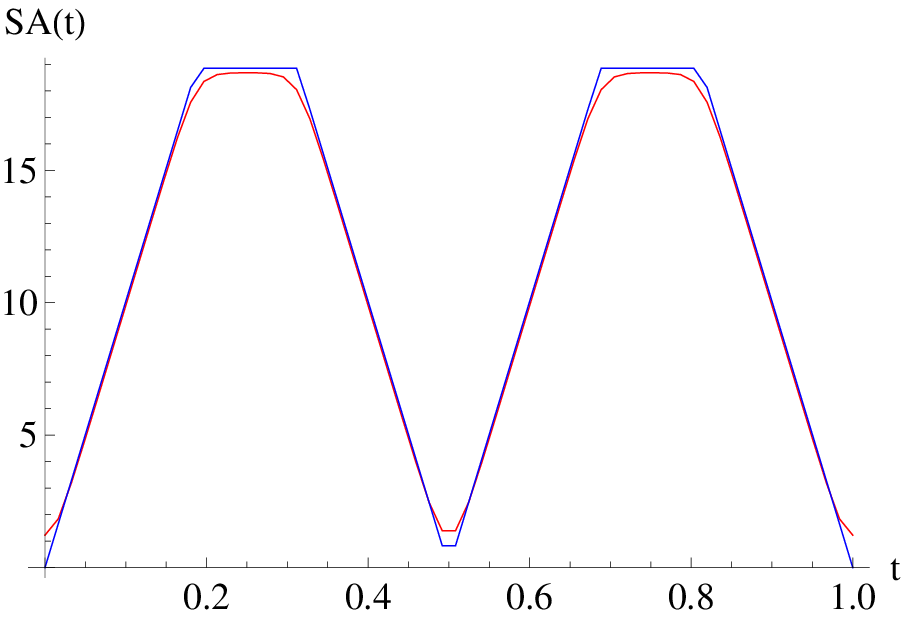}
\end{center}
\end{minipage}
\caption{Plot of the time evolution of entanglement entropy  in the quench  in the high temperature regime (blue). We also plot the evolution of the entanglement entropy expected from the quasiparticle picture (red).  
We take $L=1, \beta=\f{1}{8}, l=\f{1}{8} \;({\rm left~panel}), l=\f{3}{8}\; ({\rm right~panel}).$}
\label{fig:ee38}
\end{figure}

\begin{figure}[H]
\begin{minipage}{0.5\hsize}
\begin{center}
   \includegraphics[width=50mm]{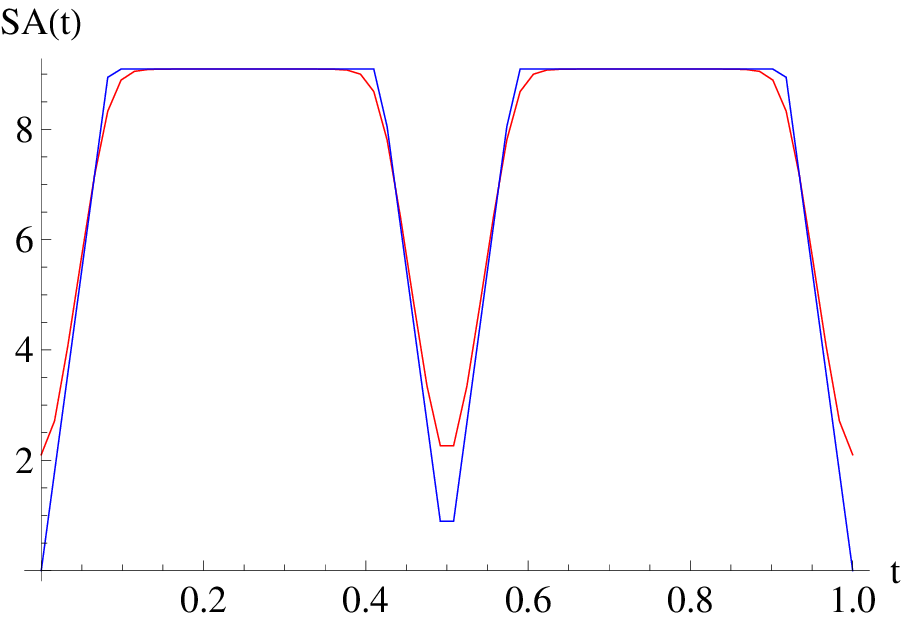}
\end{center}
\end{minipage}
\begin{minipage}{0.5\hsize}
\begin{center}
\includegraphics[width=50mm]{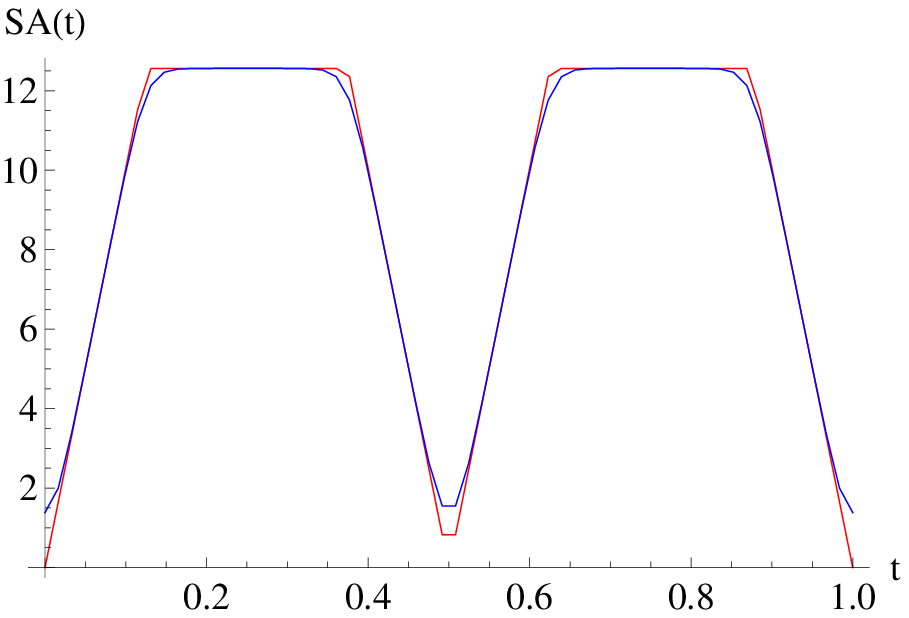}
\end{center}
\end{minipage}
\caption{Plot of the time evolution of entanglement entropy  in the quench  in the high temperature regime (blue). We also plot the evolution of the entanglement entropy expected from the quasiparticle picture (red).  
We take $L=1, \beta=\f{1}{8}, l=\f{5}{6} \;({\rm left~panel}), 
l=\f{6}{8}\; ({\rm right~panel}).$}
\label{fig:iqee}
\end{figure}

Combining everything, the total EE is 


\begin{displaymath}
S_{A}(l,t)= \f{\pi c}{6 \beta} \times \left\{
\begin{array}{l}
2t,\quad (0<t<\f{l}{2}) \quad l, \quad (\f{l}{2}<t<\f{L-l}{2}) \quad -2t+L, \quad (\f{L-l}{2}<t<\f{L}{2}) \\
2t-L \quad (\f{L}{2}<t<\f{L+l}{2}) \quad l, (\f{L+l}{2}<t<\f{2L-l}{2}), \quad -2t+2L, (\f{2L-l}{2}<t<L)
\end{array}
\right.
\end{displaymath}

As we can see, this behavior of the entanglement entropy is very
similar to that of global quenches for infinite spatial size
\eq{eq:glo}. This is natural because it takes a while until the
reflections off the walls start affecting the entanglement entropy
when the size of the subsystem $A$ is small.  As a result the
entanglement entropy is thermalized when $\frac{l}{2} \leq t \leq
\f{L-l}{2}$.  This is expected from the analysis of
\cite{Cardy:2014rqa}.  When $t \geq\f{L-l}{2}$, it starts decreasing,
leading to the `revival' behavior discussed in this paper.

\section{Conclusion}

In this paper we have explored the properties of single-interval
entanglement entropy (EE) in the presence of spatial boundaries. The
distinguishing feature arising from such boundaries is the appearance
of time-periodicity of the observables, arising from a bulk geometry
which a quotient of the AdS-Poincar\'e spacetime. The periodicity of
the entanglement entropy can be understood in terms of a periodic
motion of geodesic end-points. Another feature, for quench states
corresponding\footnote{\label{ftnt:beta-a}In the sense of footnote
  \ref{ftnt:beta}.} to high temperatures, is the appearance of
thermalization with universal exponents, followed by a revival
\cite{Cardy:2014rqa, Kuns:2014zka}. For quenches corresponding to
intermediate or low temperatures, there is an apparent puzzle: the CFT
EE depends on a four-point function which does not factorize and is
hence non-universal; the holographic result however does not depend on
details of the CFT except for the central charge $c$. We apply recent
large-$c$ techniques to resolve this puzzle.  The large $c$ limit of
the CFT EE becomes universal and exactly matches the holographic
computation at all temperatures. It would be interesting to see,
following \cite{Fitzpatrick:2016ive}, whether corrections to the
leading large $c$ limit are captured by subdominant saddle
points in the bulk. 

In this paper, we have particularized to quenches to critical theories
and modelled the quenched state by a Calabrese-Cardy (CC) type state
\cite{Calabrese:2004eu, Calabrese:2005in}. It would be interesting to
(a) generalize our results to more general states with higher chemical
potentials as discussed in \cite{Mandal:2015jla}, (b) explore
realistic quenches with boundaries in the limit of fast ramping speed
(some interesting issues have recently been raised in this context in
\cite{Das:2014hqa,Das:2014jna}) and (c) relate our results to the
study of spatial boundaries in non-conformal theories (see, e.g.
\cite{Igloi-TFI:2008}). The presence of spatial boundaries can be
regarded as a model for impurities or defects in a one-dimensional
lattice. It would be interesting to relate some of our results to
experimental situations.

An issue which has not been explored in detail in this paper, and is
currently under investigation, is the precise nature of the bulk
geometry found here. In particular, we would like to understand the
geometrical interpretation of the approximate thermalization of the
one-point function and EE and subsequent revival. These issues have
been commented upon in \cite{Kuns:2014zka, deBoer:2016bov}; however
the explicit bulk metric we have found should shed more light on this
issue. It is also interesting to see if the phenomenon of revival has
a bearing on the issue of stability of the AdS geometry discussed in,
e.g. \cite{Bizon:2011gg, Dias:2011ss, Buchel:2013uba}.\footnote{We
  thank Shiraz Minwalla for discussions on this issue.} We hope to
return to these issues shortly.

\section*{Acknowledgments}

We thank Justin David, Tarun Grover, Shiraz Minwalla, Mukund
Rangamani, and Tadashi Takayanagi for a number of
illuminating discussions. G.\,M.  would like to thank the lecturers
and participants of the Advanced String School in Bangalore, India in
June 2015, for many discussions.  The work of T.\,U. was supported in
part by the National Science Foundation under Grant No.\,NSF\,
PHY-25915.  The work of T.\,U. was supported by JSPS Postdoctoral
Research Fellowship for Young Scientists and in part by JSPS
Grant-in-Aid for JSPS Fellows, in the earlier stage of this work.

\appendix

\section{\label{app:map}Derivation of conformal maps}

In this appendix, we discuss in more detail the conformal
map mentioned in Section \ref{sec:map}.

\subsection{\label{app:r-to-c}Map from the 
rectangle to the complex plane (Euclidean)}

We consider a rectangle parametrized by
\begin{align}
w= x + i \tau, \bar w= x- i\tau, 
\; x\in [-\f L 2, \f L 2], \tau \in [-\f\b 4, \f\b 4]
\label{rectangle}
\end{align}
which we wish to map to the upper half plane
\begin{align}
z= z_2 + i z_1, \bar z= z_2 - i z_1,\;
z_1 \ge 0, z_2 \in \mathbb{R} 
\label{UHP}
\end{align}
As mentioned in  Section \ref{sec:map}, such a map is given 
by the Christoffel-Schwarz transformation \eq{christoffel}
\begin{align}
w(z)= A\int^{z}_{0} \f{dz}{\s{(z^2-b^2)(z^2-\f{1}{b^2})}} +B,
\end{align}
By definition this map satisfies the following conditions 
\begin{align}
w(b)&=\f{L}{2}+i \f{\beta}{4}, \qquad w(-b)=
\f{L}{2}-i \f{\beta}{4} \nonumber\\
w\left( \f{1}{b}\right)&=-\f{L}{2}+ i\f{\beta}{4}, 
\qquad w\left(-\f{1}{b}\right)=-\f{L}{2}-i\f{\beta}{4} 
\label{corners}
\end{align}
It is easy to see that $\b$ is given by \footnote{We have
used the convention \[
 \f{1}{\sqrt{(z^2-b^2)(z^2-\f{1}{b^2})}}=-\f{1}{\sqrt{(b^2-z^2)(\f{1}{b^2}-z^2)}}
\].}
\begin{align}
\b =2 i(w(-b)-w(b)=-2 i A \int^{b}_{-b} \f{dz}{\s{(z^2-b^2)(z^2-\f{1}{b^2})}}=
4 i A\ b K(b^4),
\end{align}
where $K(m)$ is the Elliptic K function
\[
K(m)= \int_0^1 \frac1{\sqrt{(1-x^2)(1- m x^2)}}
\]
Similarly 
\[
L =  w(b)- w(1/b)= iA\ b K(1-b^4)
\] 
The constant $B$ can now be determined from, say, the first of the conditions
\eq{corners}. We get $B= L/2$.  
The map,  therefore,  is given by 
\begin{align}
w(z) &=\f{i \beta}{4 K(b^4)} \;  sn^{-1}(\f{z}{b},b^4) +\f{L}{2} 
\label{map2}
\end{align}
where we have used the following result about the Jacobi $sn$ function
\[
\int^{z}_{0} \f{dz}{\s{(z^2-b^2)(z^2-\f{1}{b^2})}}
=  {sn}^{-1}(\f{z}{b},b^4)
\]
The inverse map is 
\begin{align}
z=b\; sn \left[ \f{4 K(b^4)}{i \beta} \left(w-\f{L}{2} \right),b^4 \right]
\label{inverse-map2}
\end{align}
where we can regard $b$ as determined from
\begin{align}
\frac{\b}{L}= \frac{4K(b^4)}{K(1-b^4)}
\label{b-l}
\end{align}
The antiholomorphic map reads as
\begin{align}
\bar z=b\; sn \left[- \f{4 K(b^4)}{i \beta} 
\left(\bar w-\f{L}{2} \right),b^4 \right]
\label{inverse-map2-zbar}
\end{align}

\paragraph{Periodicity properties:}

Note the periodicity properties
\begin{align}
z(w)= z(w+ 2 L),\;  z(w)= z(w+ i \b)
\label{periods}
\end{align}  
and the parity property
\begin{align}
z(w+i\b/4) = - z(w -i\b/4)
\label{parity}
\end{align}
which follow from the following properties of the Jacobi $sn$ function
\cite{Abramowitz}
\[
sn(u + 2 K(m), m)= - sn (u, m)= sn(u- 2 K(m), m), \;
sn(u + 2 i K(1-m), m)= sn(u, m)
\]

\paragraph{Map from $\mathbb{T}^2\to \mathbb{C}$:}

The above periodicity properties \eq{periods} imply that the
map $w \to z(w)$ can be viewed from a complex torus to the
complex plane:
\begin{align}
&\mathbb{T}^2 \ni w \mapsto z(w) = b\; sn \left( 
\f{4 K(b^4)}{i \beta} (w-\f{L}{2}), b^4 \right) \in \mathbb{C}, 
\nonumber \\
& w  \equiv x + i\tau,\, x\in [-\f L 2, \f{3L} 2], \tau\in [-
\f\b 4,
\f{3\b} 4],\;\;\; z= z_2 + i z_1,\, z_1, z_2\in \mathbb{R} 
\label{t-to-c}
\end{align}
Here the torus is represented as a rectangle (whose opposite
sides are to be identified). The
earlier rectangle \eq{rectangle} is the bottom left
`quadrant' of this larger rectangle.

We should note here that the rectangle \eq{rectangle} (or rather the
Lorentzian continuation of that) is the appropriate geometry for the
quench problem, whereas the torus described above is the appropriate
geometry for a thermal problem (where we have a periodicity
in the imaginary time).

\paragraph{Large $L$ limit:}

To see this limit, we note the following
small $m$ behaviour \cite{Abramowitz}
\begin{align}
K(1- m) \approx  \f12 \ln(16/m) + O(m), K(m)= \pi/2 + O(m)
\label{small-b}
\end{align}
Thus, the large $L$ limit corresponds to  small $b$, with
\[
\frac{\b}{L}= \frac{4K(b^4)}{K(1-b^4)} \approx \f{\pi}{\ln(2/b)}
+ O(b^4)
\] 
Using \eq{small-b} and 
\begin{align}
sn(u,m) &=  \sin(u) -\f14 m(u- \sin u\, \cos u) \cos u +
O(m^2), \nonumber\\ 
\sin(u/i) &=  -i \sinh(u)
\label{sn-sh}
\end{align}
we get from \eq{inverse-map2}
\begin{align}
z \approx -ib \sinh \left(\f{2\pi w}{\beta} - \ln(2/b)\right)
\approx -ib (-\f12) \exp\left(-\f{2\pi w}{\beta} + \ln(2/b)\right)
\approx i e^{-2\pi w/\b}
\label{high-limit-app}
\end{align}
Hence, in this limit, we recover the map \eq{high-limit} 
from the cylinder to the
complex plane with periodicity $w\equiv w+i\beta$. This periodicity is
consistent with the \eq{periods} described above, in the limit where
the aspect ratio $\b/L \to 0$, which we can regard as an infinitely
wide rectangle with width $\b$ whose opposite sides are identified
(hence, a cylinder).


\paragraph{Low temperature limit:}

Using \eq{small-b}, and \eq{b-l}, we can find the following $b\to 1-
\epsilon$ behaviour
\[
\f\b L \approx  \f 4\pi  \ln\left(\f{16}{1- b^4}\right)
\]
which shows that $b \to 1-\epsilon$ corresponds to the low temperature
$\b/L \to \infty$ limit. Using the formula
\[
sn(i u,1-m)= i \tan(u) + O(m)
\]
we can then easily derive the following low temperature limit
of \eq{inverse-map2}:
\begin{align}
z(w)=i \cot \left(\frac\pi 4 + \f{\pi w}{2 L}\right)
\label{low-limit-app}
\end{align}
This shows a periodicity $w \equiv w+ 2L$, and is
consistent with \eq{periods}. The rectangle representing
the torus \eq{t-to-c} becomes in this limit 
an infinitely high strip of width $L$ (with opposite
sides identified, hence leading to a cylinder).

\subsection{\label{zplus-minus}Lorentzian map}

As we saw in the text, the map \eq{inverse-map2} described above helps
convert observables computed in the UHP \eq{UHP} to those in the
Euclidean rectangle \eq{rectangle}. As we indicated above, the same
map also converts the complex plane to the torus \eq{t-to-c}.  As
described in Section \ref{sec:one-pt}, for real-time observables we
are interested in a Lorentzian CFT on the manifold \eq{lorentz-man}
\begin{align}
M_L= {\mathbb{I}} \times R \ni  x_\pm= x \mp t, \; x\in 
\mathbb{I} = [-L/2, L/2],
t\in \mathbb{R}
\label{lorentz-man-app}
\end{align}
which describes the Wick-rotated rectangle \eq{rectangle} (it is
an infinite strip of width $L$).  We obtain
them by analytically continuing CFT observables on the rectangle
according to \eq{analytic}:
\begin{align}
\tau= it, \to 
w= x+ i\tau= x-t\equiv x_+, \bar w= x - i\tau = x+t \equiv
x_-
\label{analytic-app}
\end{align}

For the holographic calculations described in Section \ref{sec:bulk},
we need to {\em also} analytically continue the Euclidean $z, \bar z$
plane to a Lorentzian plane $z_+, z_- \in \mathbb{R}^2$. Thus, we need
an analytic continuation of the Euclidean map
\eq{inverse-map2}, \eq{inverse-map2-zbar}. To do this, we note the
identity \cite{Abramowitz}
\begin{align}
sn (x/i, m)= -i\f{sn (x, 1-m)}{cn(x, 1-m)}
\label{identity}
\end{align}
Since under the analytic continuation \eq{analytic-app}, both $w, \bar
w$ become real, the above identity implies that both $z$ and $\bar z$,
become purely imaginary. We, therefore, define the following
analytic continuation of the complex $z$-plane to the real plane
\begin{align}
& z=i z_+,\; \bar z= - i z_-,\nonumber\\ 
& z_\pm = \left\{ \begin{array}{l} 
-i z \equiv -i(z_2+i z_1)=z_1 - z_0, \\
+ i \bar z \equiv +i(z_2-i z_1)=z_1 + z_0,
\end{array}\right. \quad  z_0\equiv i z_2
\label{analytic-z}
\end{align}
where, $z_\pm$ are given by the following functions (using
\eq{identity}) 
\begin{align}
z_\pm =  f_\pm(x_\pm),\; f_+(x) = f_-(x)= f(x) \equiv   
- b\ \f{sn\left[\f{4K(b^4)}{\beta}(x-L/2), 1-b^4\right]}{cn
\left[\f{4K(b^4)}{\beta}(x-L/2),1- b^4\right]} 
\label{fp-fm-app}
\end{align}
Note that in this map, $x_\pm= x\mp t$, with $x\in \mathbb{I}=[-L/2,
  L/2]$, $t \in \mathbb{R}$, whereas $z_\pm = z_1 \mp z_0$, $z_{0,1}
\in \mathbb{R}$. Thus the above map is a map from the strip $M_L$
\eq{lorentz-man} to the real plane $\mathbb{R}^2$. The map is clearly
conformal, which satisfies the property $ dz_+ dz_- = f'(x_+) f'(x_-)
dx_+ dx_- $. Note that the Lorentzian version of the high temperature map
\eq{high-limit-app} becomes the Rindler map:
\begin{align}
z_\pm =  \exp(-\f{2\pi x_\pm}{\b})
\label{rindler}
\end{align}

\subsection{A different Euclidean map}

We note that the map \eq{inverse-map2} is not the unique
one that maps the rectangle to the UHP. We may consider, e.g.,
a different assignment of the corners to the boundary of
the UHP:
\begin{align}
w\left(-\f{1}{b}\right)&=\f{L_1}{2}-\f{i\beta}{4}, 
\quad w\left(\f{1}{b}\right)=-\f{L}{2}-\f{i\beta}{4} \nonumber \\
w(-b)&=\f{L}{2}+\f{i\beta}{4}, 
\quad w(b)=-\f{L}{2}+\f{i\beta}{4} 
\end{align}
By using methods similar to Section \ref{app:r-to-c}, we arrive at 
the following map
\begin{equation}
z=-b \;{\rm sn} \left[ \f{2K(b^4)}{L} (w-\f{i\beta}{4}), b^4\right] 
\label{inverse-map1}
\end{equation}
where $b$ now is given by
\[
\f{\beta}{L} =\f{K(1-b^4)}{K(b^4)}.
\]

\begin{figure}[H]
\begin{center}
\kern-20pt\includegraphics[width=8cm,height=4.5cm]{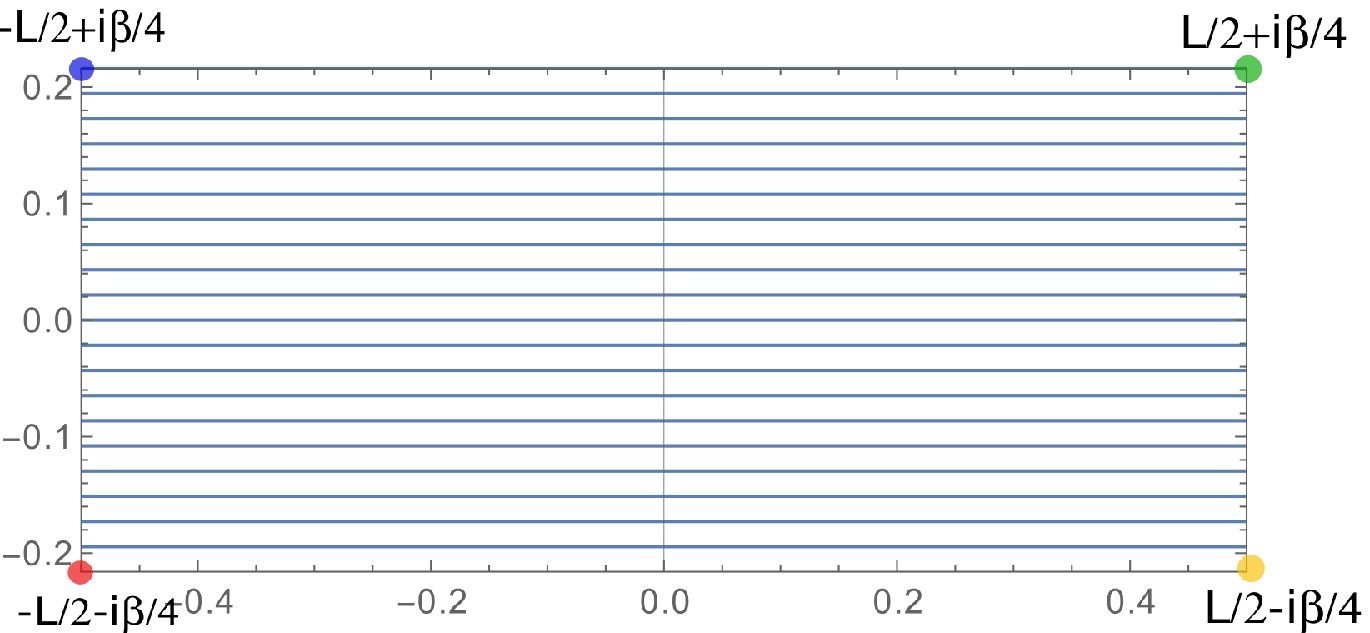}
\kern+20pt\includegraphics[width=7cm,height=7cm]{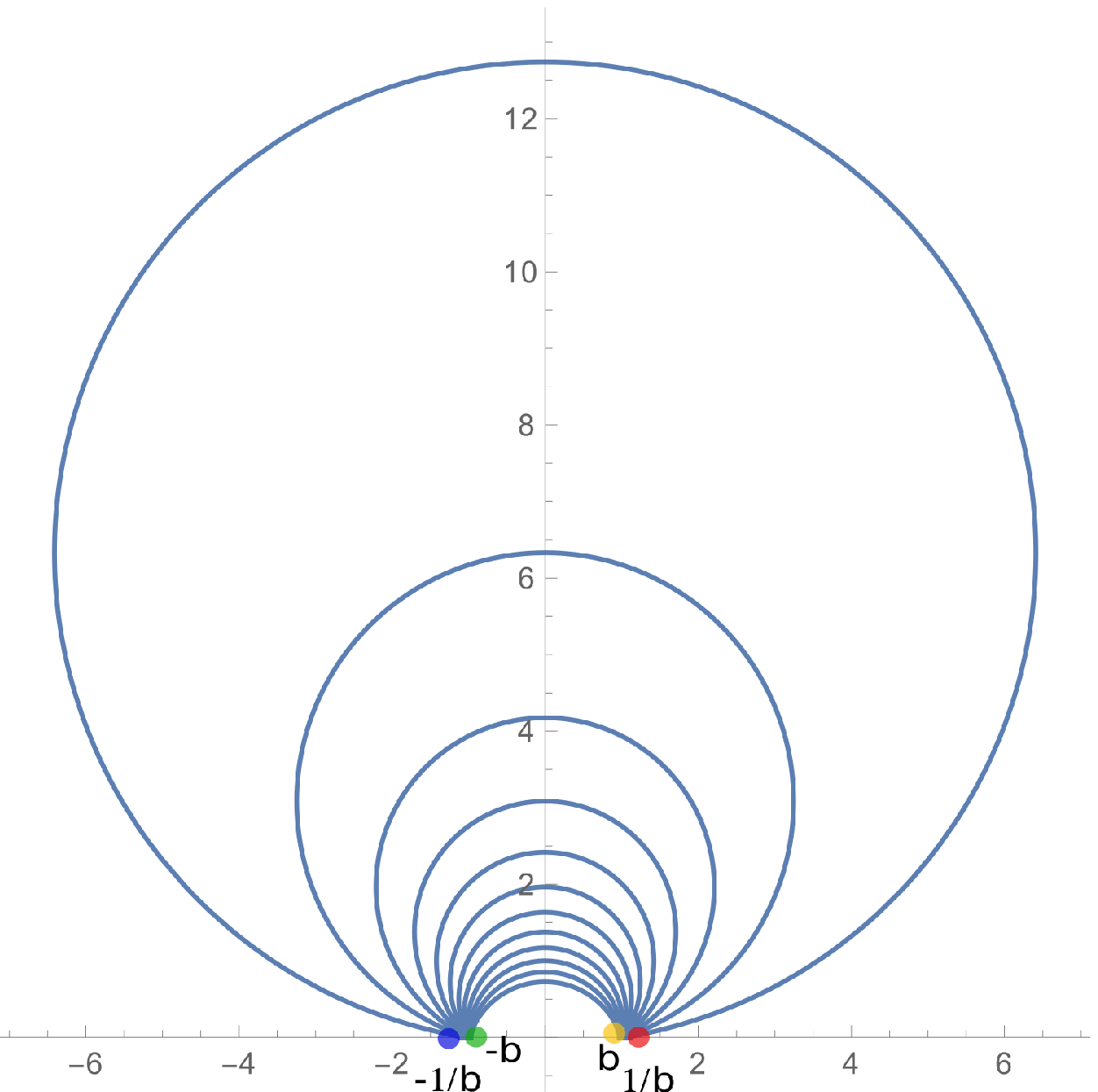}
\end{center}
\caption{\footnotesize The second map between the upper half plane and
  the rectangle. The colour coding represents the mapping of the
  corners to the boundary of the UHP. See (\ref{inverse-map2}) and
  (\ref{map2}). The time evolution contours in the rectangle are
  mapped to the UHP as shown on the right. We have chosen
  $L=\beta=1$. It is clear that this second map is roughly obtained
  from the first map by exchanging the horizontal and vertical axes of
  the complex plane.}
\label{fig:map2}
\end{figure}

The periodicity property of this map is the same as in \eq{periods}.
This map has the property that its low temperature limit is the standard
cylinder map (with periodicity $2L$), as 
\begin{align}
z\approx i \exp(i\f{\pi w}{L})
\label{low-limit-2}
\end{align}
The Euclidean map \eq{inverse-map1} 
cannot be analytically continued to a Lorentzian map 
as in \eq{analytic}. Here the continuation is $w= i x_+,
\bar w= - i x_-$. This means that in the Euclidean theory 
we need to call $w=: \tau + i x,
\bar w= \tau - ix$, and then analytically continue $\tau= it$
(here $x_\pm = x\pm t$). The accompanying continuation of the
$z$-plane is $z= i z_+, \bar z= -i z_-$. As a simple check, note
that \eq{low-limit-2} now becomes a real conformal map 
\[
z_\pm
=  \exp(-\f{\pi x_\pm}{L})
\]
Comparing with \eq{rindler}, we can see that this 
second map  is indeed related to the first map  by
an exchange of the horizontal and vertical axes, along with
$\b \leftrightarrow  2L$ (also, compare Figures 
\ref{fig:map} and \ref{fig:map2}).

\section{\label{app:one-pt}Decay rate in the high temperature limit}

We will consider a one point function at the point $x=0$ and at time
$t$ in the high temperature limit, where
$\f{L}{\beta}\gg\f{t}{\beta}\gg 1$. Using the high temperature
Lorentzian map \eq{rindler}, we find that the Lorentzian continuation
of the connected correlator $\lan ... \ran_{\mathbb{C}}$ in
\eq{connected-2pt-primary} gives
\begin{align}
 \left(\frac{(z_+-z_-)}{b}\right)^{-2h}
 \simeq 2^{2h}\exp(-\f{2\pi}{\beta}(2h)(t+\f{L}{2}))
\end{align}
The Jacobian terms in \eq{connected-2pt-primary} become
\begin{align}
 \left(\frac{z_+'(x_+)z_-'(x_-)}{b}\right)^h
\simeq (\f{2\pi}{\beta})^{2h}2^{-2h}\left(\exp(\f{4\pi t}{\beta})+
\exp(\f{2\pi L}{\beta})\right)^h
\end{align}
In the limit $\f{L}{2\beta}\gg\f{t}{\beta}$, 
the second term in the above expression dominates over the first. Combining the above two expressions,
we get, 
\begin{align}
 \langle\phi(w,\bar w)\rangle_{\text{rect}}
\simeq \left(\f{2 \pi}{\beta}\right)^{2h}\exp\left(-\f{4\pi h}{\beta} t
\right)
\end{align}
Thus, the decay rate of the one point function in the limit
$\f{L}{\beta}\gg\f{t}{\beta}\gg 1$ matches exactly with the
decay rate obtained in \cite{Cardy:2014rqa} and \cite{Mandal:2015jla}
in the case of an infinite strip ($L\rightarrow\infty$).

\section{The BTZ and the global AdS metric}\label{app:poincare-btz}

We will work out the metric \eq{banados} in case the maps
$f_\pm(x_\pm)$ are given by the non-compact limit $L \to \infty$ limit
\eq{high-limit} and the low temperature $\b \to \infty$ limits
\eq{low-limit} of the general transformation
\eq{inv-map2}.\footnote{\label{ftnt:limits} We must distinguish
  between (a) the non-compact limit $L \to \infty$ and (b) the high
  temperature limit $\b \to 0$, although in both cases the aspect
  ratio $L/\b \to \infty$. The difference is that in (a), ratios such
  as $\b/L, x_\pm/L, l/L \to 0$ while ratios such as $x_\pm/\b, l/\b$
  are not scaled, whereas in (b) ratios such $L/\b, x_\pm/\b, l/\b \to
  \infty$, while $x_\pm/L, l/L$ are not scaled. In other words, in (a)
  only $L \to \infty$ and all other length scales are held fixed,
  while in (b) only $\b \to 0$ and all other length scales are held
  fixed. Similarly limits of (c) low temperature $\b \to \infty$
  (where $L, x_\pm, l$ are held fixed) and (d) small system size $L
  \to 0$ (where $\b, x_\pm, l$ are held fixed), are different. In
this section we consider the limits (a) and (c).}

\paragraph{Large $L$: BTZ}

Using the map \eq{high-limit} we first find that
\begin{align}
L_+ = L_- \equiv \bl =   4\pi^2/\b^2
\label{energy-btz}
\end{align}
which is consistent with \eq{t-vs-l} and \eq{thermal}. This allows
us to write the map \eq{high-limit} as
\begin{align}
z_\pm= f_\pm(x_\pm) = \exp[-\sqrt{\bl} x_\pm]
\label{carlip}
\end{align}
The denominator $D$ in \eq{roberts} becomes
\begin{align}
D=  \bl \left(\bl z^2+4\right)
   e^{-2\sqrt{\bl} x},
\label{denom}
\end{align}
The `large diffeomorphism' \eq{roberts} in this case is,
therefore, given by
\begin{align}
z_+= f(x_+) \equiv
\frac{e^{-\sqrt{\bl} x_+} \left(4-\bl z^2\right)}{\bl z^2+4},
\;
z_-=f(x_)
\;
\zeta= \frac{4 z \sqrt{\bl e^{-\sqrt{\bl} (x_-+x_+)}}}{\bl z^2+4}
\label{roberts-btz}
\end{align}
The resulting metric \eq{banados}, turns out to be
(see \cite{Roberts:2012aq, Ugajin:2013xxa, Mandal:2014wfa} for more
details) 
\begin{align}
ds^2=\f{dz^2}{z^2}   -\frac{\left(\bl z^2-4\right)^2}{16 z^2} {dt}^2
+ \frac{\left(\bl z^2+4\right)^2}{16 z^2} {dx}^2, 
\label{btz}
\end{align}
which represents a BTZ black hole with horizon at 
\begin{align}
z_h = 2/\sqrt{\bl}
\label{btz-horizon}
\end{align}
In the large $L$ limit the size
of the spatial cycle effectively becomes decompactified;
thus the above metric represents a BTZ black string. 
Thus, the BTZ black string is the bulk dual of the CFT on the
(analytically continued) cylinder, which, of course, is the
Lorentzian equivalent of the statement that the Euclidean black
string is the bulk dual of the thermal CFT which is represented
by the Euclidean cylinder.

\paragraph{The low temperature limit: Global AdS}

In the limit $\b \gg L$, we apply the map \eq{low-limit} to
the transformation \eq{roberts}. Here we find that
\begin{align}
L_+ = L_- \equiv \bl = - \pi^2/L^2
\label{energy-soliton}
\end{align}
which is consistent with \eq{t-vs-l} and \eq{casimir}.

By going through similar steps as the above, we find the 
final form of the metric \eq{banados} as 
\begin{align}
ds^2 =\f{dz^2}{z^2} 
-\frac{\left(\bl z^2+4\right)^2}{16 z^2}
 dt^2 + \frac{\left(\bl z^2-4\right)^2}{16 z^2} dx^2 
\label{ads-soliton}
\end{align}
which represents global AdS, since it is the spatial cycle
$x$ which shrinks to zero size at $z=2/\sqrt{\bl}$.  

\bibliographystyle{jhepmod}
\bibliography{bq}

\end{document}